\documentclass[aps,twocolumn,showpacs,floatfix,amssymb]{revtex4}
\usepackage{graphicx}
\begin{document}

\title{Kinematic spin-fluctuation mechanism of
high-temperature superconductivity}
\author{Nikolay M. Plakida$^{1}$ and Viktor S. Oudovenko$^{2}$}
 \affiliation{$^1$Joint Institute for Nuclear Research,
 141980 Dubna, Russia}
  \affiliation{$^2$Rutgers University,  New Jersey 08854, USA }

\date{\today}

\begin{abstract}

We study $d$-wave superconductivity  in the extended Hubbard
model in the strong correlation limit for a large intersite
Coulomb repulsion $V$.  We argue that in the Mott-Hubbard regime
with two Hubbard subbands there emerges a new energy scale for the
spin-fluctuation coupling of electrons  of the order  of the
electronic kinetic energy $ W $ much larger than the exchange
energy $J$. This coupling is induced by the kinematic interaction
for the Hubbard operators  which results in the kinematic
spin-fluctuation pairing mechanism for $\,V \lesssim W $. The
theory is based on the Mori projection technique in the equation
of motion method for the Green functions in terms of the Hubbard
operators. The doping dependence of superconductivity
temperature  $T_c$ is calculated for various values of $U$ and
$V$.
\end{abstract}


 \maketitle

\section{Introduction}
\label{sec:1}

One of  crucial issues in the superconductivity  theory is to
disclose  the mechanism of high-temperature superconductivity
(HTSC) in  cuprates  (see, e.g.~\cite{Schrieffer07,Plakida10}).
In  early studies of the problem,  a model of strongly correlated
electrons was proposed by Anderson~\cite{Anderson87} where
superconductivity occurs at finite doping in the resonating
valence bond state (RVB) due to the antiferromagnetic (AF)
superexchange interaction $J$. However, the intersite Coulomb
interaction (CI)  $V$ that  in cuprates is of the order of $J$
may destroy the RVB state and superconducting pairing. Recently a
competition of the intersite CI $V$ and pairing induced  by the
on-site CI $U$ in the Hubbard model~\cite{Hubbard63}  or by the
intersite CI $V$ was actively discussed. In particular, in
Ref.~\cite{Alexandrov11} it was stressed that a contribution from
the repulsive well-screened weak CI in the first order strongly
suppresses the pairing induced by contributions of higher orders,
and a possibility for superconductivity ``from repulsion'' was
questioned. Using the renormalization group method in
Ref.~\cite{Raghu12}   the extended Hubbard model with CI $V$ was
studied where superconducting pairing of various symmetries,
extended $s$-, $p$-, and $d$-wave types was found depending on
the electron concentration and   $V$. Following the original idea
of Kohn-Luttinger~\cite{Kohn65}, in Ref.~\cite{KaganM11} it was
shown that the $p$-wave superconductivity exists  in the
electronic gas at low density with a strong repulsion $U$ and a
relatively strong intersite  CI $V$ (see, also~\cite{Efremov00}
and references therein). Studies of the phase diagram within the
extended Hubbard model in the weak correlation limit  have shown
that superconducting pairing of different types of symmetry, $s$,
$p$,  $d_{xy}$, and $d_{x^2 -y^2}$ can occur depending on the CI
between the nearest $V_1$ and next $V_2$ neighbor sites  and
electron hopping parameters between distant sites in a broad
region of electron concentration~\cite{KaganM13}.
\par
However, in these investigations the Fermi-liquid model in the
weak correlation limit, $U \lesssim W $, was considered, while
cuprates are the Mott-Hubbard (more accurately, charge-transfer)
doped insulators where  a theory of strongly correlated electronic
systems should be applied  for $U \gtrsim W $.  Here  $W \sim
4\,t$ is the electronic kinetic energy  for the two-dimensional
Hubbard model with the nearest neighbor hopping parameter $t$. In
the limit of strong correlations various numerical methods for
finite clusters are commonly used. There are many investigations
of the conventional Hubbard model (see,
e.g.~\cite{Dagotto94,Bulut02,Scalapino07,Senechal12r}) but only
few studies of the extended Hubbard model in which the intersite
CI $V$ is taken into account. In particular, in
Refs.~\cite{Plekhanov03,Senechal12,Raghu12a} the extended Hubbard
model was considered in a broad region of $U$ and $V$. The
results of Refs.~\cite{Plekhanov03,Senechal12} show that a
strong  on-site repulsion $U$ effectively enhances the $d$-wave
pairing which is preserved  for large values of  $V \gg J $. In
Ref.~\cite{Raghu12a} using the slave-boson representation it was
found that superconductivity is destroyed at a small value of $V
= J$. We discuss these results in more detail in
Sec.~\ref{sec:10} by comparing them with our findings.
\par
In our recent paper~\cite{Plakida13} we studied the extended
Hubbard model in the limit of strong correlations  by taking into
account  the CI $V$ and  electron-phonon coupling. It was found
that $d$-wave pairing with high-$T_c$ is mediated by the strong
kinematic interaction of electrons with spin fluctuations.
Contributions coming from weak  CI $V$ and phonons turned out to
be small since only $\, l=2\,$ harmonics of the interactions give
a contribution to the $d$-wave pairing.
\par
In the this paper, we consider superconductivity in the
two-dimensional extended Hubbard model  with a large  intersite
Coulomb repulsion $V$  in the limit of strong correlations  to
elucidate   the spin-fluctuation   mechanism of high-temperature
superconductivity. We argue that in the two-subband regime for
the Hubbard model for $U \gtrsim 6\, t$ a spin-electron kinematic
interaction is evolved from complicated commutation relations for
the Hubbard operators (HOs)~\cite{Hubbard65}.  This interaction
brings about the weak exchange interaction $J = 4t^2/U$ due to
interband hopping and at the same time  intraband hopping results
in a much stronger  kinematic interaction $\,g_{sf} \sim W \gg J$
of electrons with spin excitations.  Therefore, the exchange
interaction $J$ is not so important for the spin-fluctuation
pairing driven by the strong kinematic interaction $\,g_{sf} $.
We calculate the doping dependence of superconducting $T_c$ for
various values of $U$ and $V$ and show that as long as $V$ does
not exceed the kinematic interaction, $V \lesssim W $, the
$d$-wave pairing is preserved. In calculations we use the
Mori-type projection technique~\cite{Mori65} in the equation of
motion method for thermodynamic Green functions
(GFs)~\cite{Zubarev60} expressed in terms of the HOs. The
self-energy in the Dyson equation is calculated in the
self-consistent Born approximation (SCBA) as in our previous
publications~\cite{Plakida07,Plakida13}.
\par
In Sec.~\ref{sec:2} the two-subband extended Hubbard model is
introduced and equations for the GFs in the Nambu representation
are derived.   A self-consistent system of equations for GFs and
the self-energy is formulated in Sec.~\ref{sec:4}. Results and
discussion are presented  in Sec.~\ref{sec:7}. Concluding remarks
are given in Sec.~\ref{sec:11}.

\section{General formulation}
\label{sec:2}

\subsection{Extended Hubbard model}
\label{sec:2a}

We consider the extended  Hubbard model  on a square lattice
\begin{eqnarray}
H &= &\sum_{i \neq j, \sigma} \, t_{ij} \, a_{i\sigma}^{\dag}
a_{j\sigma} - \mu \, \sum_{i} N_{i}
\nonumber \\
& + & (U/2) \,\sum_{i} N_{i \sigma}  N_{i \bar\sigma} +
(1/2)\,\sum_{ i \neq j} V_{ij}\, N_{i} N_{j},
 \label{1}
\end{eqnarray}
where $t_{i,j}$ is the single-electron hopping parameters,
$a^{\dag}_{i\sigma}$ and $a_{i\sigma}$ are the Fermi creation and
annihilation  operators for  electrons with spin $\sigma/2
\;(\sigma = \pm 1 = (\uparrow, \downarrow), \, \bar\sigma =
-\sigma) $ on the lattice site $i$, $U$ is the on-site CI and the
$V_{ij}$ is the intersite CI. $ \, N_{i} =\sum_{\sigma}
N_{i\sigma}, \, N_{i\sigma}=a^{\dag}_{i\sigma}a_{i\sigma} $ is
the number operator and $\mu$ is the chemical potential.

In the strong correlation limit the model describes the
Mott-Hubbard insulating state  at half-filling ($n = \langle
N_{i}  \rangle = 1$) when the conduction band splits into two
Hubbard subbands. In this case the  Fermi operators
$a^\dag_{i\sigma}, \, a_{i\sigma}$ in (\ref{1}) fail to describe
single-particle electron excitations in the system and the
Fermi-liquid picture becomes inadequate for cuprates. The
projected-type operators, the Hubbard operators (HOs), referring
to the two subbands, singly occupied $a^\dag_{i\sigma}(1- N_{i
\bar\sigma})$ and doubly occupied $a^\dag_{i\sigma} N_{i
\bar\sigma}$, must  be introduced. In terms of the HOs the model
(\ref{1}) reads
\begin{eqnarray}
 H &= & \varepsilon_1\sum_{i,\sigma}X_{i}^{\sigma \sigma}
  + \varepsilon_2\sum_{i}X_{i}^{22}
+  \frac{1}{2} \sum_{i\neq j}\,V_{ij} N_i N_j
\nonumber \\
& + &\sum_{i\neq j,\sigma}\, t_{ij}\,\bigl\{ X_{i}^{\sigma
0}X_{j}^{0\sigma}
  + X_{i}^{2 \sigma}X_{j}^{\sigma 2}
 \nonumber \\
& + & \sigma \,(X_{i}^{2\bar\sigma}X_{j}^{0 \sigma} + {\rm
H.c.})\bigr\},
 \label{2}
\end{eqnarray}
where $\varepsilon_1 = - \mu$ is the single-particle energy  and
$\varepsilon_2 =  U - 2 \mu $ is the two-particle energy. The
matrix HOs $X_{i}^{\alpha\beta} = |i\alpha\rangle\langle i\beta|$
describes  transition from the state $|i,\beta\rangle$ to the
state $|i,\alpha\rangle$ on a lattice site $i$ taking into
account four  possible states for holes: an empty state $(\alpha,
\beta =0) $, a singly occupied hole state $(\alpha, \beta =
\sigma)$, and a doubly occupied hole state $(\alpha, \beta = 2)
$. The number operator and the spin operators in terms of the HO
are defined as
\begin{eqnarray}
  N_i &=& \sum_{\sigma} X_{i}^{\sigma \sigma} + 2 X_{i}^{22},
\label{3a}\\
S_{i}^{\sigma} & = & X_{i}^{\sigma\bar\sigma} ,\quad
 S_{i}^{z} =  (\sigma/2) \,[ X_{i}^{\sigma \sigma}  -
  X_{i}^{\bar\sigma \bar\sigma}] .
\label{5}
\end{eqnarray}
The chemical potential $\mu$ is determined from the equation for
an average occupation number for holes
\begin{equation}
  n =  1 + \delta = \langle \, N_i \rangle ,
    \label{4}
\end{equation}
where  $\langle \ldots \rangle$ denotes the statistical average
with the Hamiltonian (\ref{2}).
\par
The HOs obey the completeness relation   $\, X_{i}^{00} +
 \sum_{\sigma} X_{i}^{\sigma\sigma}  + X_{i}^{22} = 1 $ which
rigorously preserves the constraint  that only one  quantum state
$\alpha$ can be occupied on any lattice site $i$.  The
commutation relations for the HOs
\begin{equation}
\left[X_{i}^{\alpha\beta}, X_{j}^{\gamma\delta}\right]_{\pm}=
\delta_{ij}\left(\delta_{\beta\gamma}X_{i}^{\alpha\delta}\pm
\delta_{\delta\alpha}X_{i}^{\gamma\beta}\right)\, ,
 \label{6}
\end{equation}
with the upper  sign  for the Fermi-type operators (such as
$X_{i}^{0\sigma}$) and the lower sign for the Bose-type operators
(such as   $N_i$  (\ref{3a}) or the spin operators (\ref{5}))
result in the so-called {\it kinematic}  interaction. To
demonstrate this let us consider the equation of motion for the
HO $\, X\sb{i}\sp{\sigma 2} =
a^{\dag}_{i\sigma}a_{i\sigma}a_{i\bar\sigma}\, $ in the
Heisenberg representation $(\hbar = 1)$:
\begin{eqnarray}
 i\frac{d}{d t}  X\sb{i}\sp{\sigma 2} &= &[X\sb{i}\sp{\sigma 2}, H] =
   (U - \mu + \sum\sb{l}  V_{i l}\,N_{l} )\, X_{i}^{\sigma 2}\,
\nonumber \\
  &+& \sum\sb{l,\sigma '}t\sb{il} \left(
    B\sb{i\sigma\sigma '}\sp{22} X\sb{l}\sp{\sigma ' 2} -
    \sigma \, B\sb{i\sigma\sigma '}\sp{21}
    X\sb{l}\sp{0\bar\sigma '} \right)
\nonumber \\
 &-& \sum\sb{l} t\sb{il}\, X\sb{i}\sp{02}  \left(
    X\sb{l}\sp{\sigma0} +  \sigma
    X\sb{l}\sp{2 \bar\sigma} \right),
 \label{7}
\end{eqnarray}
Here $B\sb{i\sigma\sigma'}\sp{\eta \zeta}$ are the Bose-type
operators,
\begin{eqnarray}
  B\sb{i\sigma\sigma'}\sp{22} & = & (X\sb{i}\sp{22} +
   X\sb{i}\sp{\sigma\sigma}) \, \delta\sb{\sigma'\sigma} +
   X\sb{i}\sp{\sigma\bar\sigma} \, \delta\sb{\sigma'\bar\sigma}
\label{8a} \\
  &  = &( N\sb{i}/2 +  \sigma\, S\sb{i}\sp{z}) \, \delta\sb{\sigma'\sigma} +
    S\sb{i}\sp{\sigma} \, \delta\sb{\sigma'\bar\sigma},
\nonumber \\
  B\sb{i\sigma\sigma'}\sp{21} & = & ( N\sb{i}/2 +
   \sigma S\sb{i}\sp{z}) \, \delta\sb{\sigma'\sigma} -
   S\sb{i}\sp{\sigma}\,  \delta\sb{\sigma'\bar\sigma}.
  \label{8b}
\end{eqnarray}
We see that the hopping amplitudes depend on the number operator
(\ref{3a}) and the spin operators (\ref{5}) which results in the
kinematic interaction describing  effective scattering of
electrons on spin and charge fluctuations. In  phenomenological
models for cuprates  a dynamical coupling of electrons with spin
and charge fluctuations is introduced specified by fitting
parameters, while in Eq.~(\ref{7}) the interaction is determined
by the hopping energy $t_{ij}$ fixed by the electronic dispersion.

\subsection{Green functions}
\label{sec:3}

To consider  superconducting pairing in the  model (\ref{2}), we
introduce the two-time thermodynamic  GF~\cite{Zubarev60}
expressed in terms of the four-component Nambu operators, $\,
\hat X_{i\sigma}$ and   $\, \hat
X_{i\sigma}^{\dagger}=(X_{i}^{2\sigma}\,\, X_{i}^{\bar\sigma
0}\,\, X_{i}^{\bar\sigma 2}\,\, X_{i}^{0\sigma}) \,$:
\begin{eqnarray}
 {\sf G}_{ij\sigma}(t-t') & = & -i \theta(t-t')\langle \{
 \hat X_{i\sigma}(t) ,  \hat X_{j\sigma}^{\dagger}(t')\}\rangle
 \nonumber \\
 & \equiv & \langle \!\langle \hat X_{i\sigma}(t) \mid
    \hat X_{j\sigma}^{\dagger}(t')\rangle \!\rangle,
 \label{9}
\end{eqnarray}
where $ \{A, B\} = AB + BA$,  $ A(t)= \exp (i Ht) A\exp (-i Ht)$,
and $\theta(x) = 1 $ for $x > 0 $ and $\theta(x) = 0 $ for $x < 0
$. The Fourier representation in $({\bf k}, \omega) $-space is
defined by the relations:
\begin{eqnarray}
{\sf G}_{ij\sigma}(t-t') &= &
\frac{1}{2\pi}\int_{-\infty}^{\infty} dt e^{- i(t-t')} {\sf
G}_{ij\sigma}(\omega),
 \label{9ft}\\
{\sf G}_{ij\sigma}(\omega) & = &\frac{1}{N}\,\sum_{\bf
k}\exp[i{\bf k (i-j)}] {\sf G}_{\sigma}({\bf k}, \omega).
    \label{9fk}
\end{eqnarray}
The GF (\ref{9fk}) is convenient to write in the matrix form
\begin{equation}
{\sf G}_{\sigma}({\bf k}, \omega)=
  {\hat G_{\sigma}({\bf k}, \omega)  \quad \quad
 \hat F_{\sigma}({\bf k}, \omega) \choose
 \hat F_{\sigma}^{\dagger}({\bf k}, \omega) \quad
   -\hat{G}_{\bar\sigma}(-{\bf k}, -\omega)} ,
 \label{9m}
\end{equation}
where the normal $\hat G_{\sigma}({\bf k}, \omega)$ and anomalous
(pair) $\hat F_{\sigma}({\bf k}, \omega) $  GFs are
 $2\times 2$ matrices for  two Hubbard subbands:
\begin{equation}
\hat G_{\sigma}({\bf k}, \omega) = \langle\! \langle \left(
\begin{array}{c}
     X_{\bf k}^{\sigma2}  \\
     X_{\bf k}^{0 \bar\sigma } \\
        \end{array}\right)  \mid
  X_{\bf k}^{2\sigma} X_{\bf k}^{\bar\sigma 0}
 \rangle \! \rangle_{\omega},
 \label{9a}
\end{equation}
\begin{equation}
\hat F_{\sigma}({\bf k}, \omega)  = \langle\! \langle \left(
\begin{array}{c}
     X_{\bf k}^{\sigma2}  \\
     X_{\bf k}^{0 \bar\sigma } \\
        \end{array}\right)  \mid
  X_{-\bf k}^{\bar\sigma 2} X_{-\bf k}^{0 \sigma}
 \rangle \! \rangle_{\omega}.
 \label{9b}
\end{equation}
To calculate the GF (\ref{9})  we use the equation of motion
method by differentiating the  GF with respect to  time $t$ and
$t'$. As described in detail in Refs.~\cite{Plakida07,Plakida13},
using the Mori-type projection method~\cite{Mori65} we derive  an
exact representation for the GF (\ref{9m})  in the form of  the
Dyson equation
\begin{equation}
 {\sf G}\sb{\sigma}({\bf k}, \omega) =
  \left[\omega \tilde{\tau}\sb{0} - {\sf E}\sb{\sigma}({\bf k})
  -    {\sf  Q} {\sf \Sigma}_{\sigma}({\bf k}, \omega)
  \right] \sp{-1} {\sf Q},
\label{14}
\end{equation}
where $\tilde{\tau}\sb{0}$ is the $4\times 4$ unit matrix.  The
electron excitation spectrum  in the generalized mean-field
approximation (GMFA) is determined by the time-independent matrix
of correlation functions:
\begin{eqnarray}
  {\sf E}\sb{\sigma}({\bf k})&=&
  \frac{1}{N}\sum_{\bf k}\exp[i{\bf k (i-j)}] \langle \{ [\hat X\sb{i\sigma}, H],
    \hat X\sb{j\sigma}\sp{\dagger} \} \rangle {\sf Q}^{-1}
 \nonumber\\
   &= & \left(
\begin{array}{cc}
  \hat{\varepsilon}({\bf k})  & \hat{\Delta}_{\sigma}({\bf k}) \\
     \hat{\Delta}_{\sigma}^{*}({\bf k}) &
     -\hat{\varepsilon}_{\bar\sigma}({\bf k})
\end{array}\right)   ,
\label{12}
\end{eqnarray}
where $\hat{\varepsilon}({\bf k})$ and $
\hat{\Delta}_{\sigma}({\bf k})$ are the normal and anomalous
parts of the energy matrix.  The parameter ${\sf Q} = \langle
\{\hat X\sb{i\sigma},\hat X\sb{i\sigma}\sp{\dagger}\}\rangle   =
\hat{\tau}_{0} \times \hat Q\,$ where   $\, \hat{\tau}_{0}$ is
the $2 \times 2$ unit matrix and $\, \hat Q =
\left(  \begin{array}{cc} Q\sb{2} & 0 \\
      0 & Q\sb{1} \end{array}  \right)\,$ takes into account a redistribution of the
spectral weights with doping of the Hubbard subbands
 $\, Q\sb{2} = \langle X\sb{i}\sp{22} +
X\sb{i}\sp{\sigma\sigma} \rangle = n/2 \,$ and $\, Q\sb{1} =
\langle X\sb{i}\sp{00} + X\sb{i}\sp{\bar\sigma \bar\sigma}
\rangle = 1-Q\sb{2}\, $.

The self-energy operator in Eq.~(\ref{14})
\begin{equation}
 {\sf  Q} {\sf \Sigma}\sb{\sigma}({\bf k}, \omega) =
    \langle\!\langle {\hat Z}\sb{{\bf k}\sigma}\sp{(\rm ir)} \!\mid\!
     {\hat Z}\sb{{\bf k}\sigma}\sp{(\rm ir)\dagger} \rangle\!\rangle
      \sp{(\rm pp)}\sb{\omega}\;{\sf  Q}\sp{-1} ,
\label{15}
\end{equation}
determined by irreducible operators $\, \hat
Z\sb{i\sigma}\sp{(\rm ir)}= [\hat X\sb{i\sigma}, H] -
\sum\sb{l}{\sf E}\sb{il\sigma} \hat X\sb{l\sigma}\, $, describes
processes of inelastic scattering of electrons (holes) on spin
and charge fluctuations due to the kinematic interaction and  CI
$V_{ij}$ (see Eq.~(\ref{7})). The self-energy operator (\ref{15})
can be  written in the same matrix form as the  GF (\ref{9m}):
\begin{equation}
 {\sf  Q} {\sf \Sigma}\sb{\sigma}({\bf k}, \omega) =  {\hat M_{\sigma}({\bf k}, \omega)
 \quad  \quad
\hat\Phi_{\sigma}({\bf k}, \omega) \choose
\hat\Phi_{\sigma}^{\dagger} ({\bf k}, \omega)\quad
-\hat{M}_{\bar\sigma}({\bf k}, -\omega)} {\sf Q}^{-1}  \, ,
 \label{23}
\end{equation}
where the matrices $\hat M$ and $\hat\Phi$  denote the respective
normal and anomalous (pair) components of the self-energy
operator.

The system of equations for the $(4 \times 4)$ matrix GF
(\ref{9m}) and the self-energy (\ref{23}) can be reduced to a
system of equations for the normal ${\hat G}_\sigma({\bf
k},\omega)$ and the pair ${\hat F}_\sigma({\bf k},\omega)$ $(2
\times 2)$ matrix components. Using representations for the
energy matrix (\ref{12}) and the self-energy (\ref{23}), we
derive  for these components  the following system of matrix
equations:
\begin{eqnarray}
{\hat G}({\bf k},\omega) & = & \Bigl(
  \hat {G}_{N}({\bf k},\omega)^{-1}
\nonumber \\
& +  & \hat{\varphi}_\sigma({\bf k},\omega)\,
  \hat{G}_{N}({\bf k},- \omega)\,\hat{\varphi}^{*}_\sigma({\bf
k},\omega)  \Bigr)^{-1} \, \hat{Q}, \qquad
 \label{24} \\
{\hat F}_\sigma({\bf k},\omega) & = & -\hat{G}_{N}({\bf k},-
\omega)\,\hat{\varphi}_\sigma({\bf k},\omega) \,
 \hat{G}({\bf k},\omega) ,
 \label{25}
\end{eqnarray}
where we introduced  the normal state  GF
\begin{eqnarray}
{\hat G}_{N}({\bf k},\omega)& = & \Bigl( \omega \hat\tau_0 -
\hat{\varepsilon}({\bf k}) -
  \hat{M}({\bf k},\omega)/ \hat{Q} \Bigr)^{-1},
\label{26}
 \end{eqnarray}
and the  superconducting gap function
\begin{eqnarray}
{\hat \varphi}_\sigma({\bf k},\omega)& = &
\hat{\Delta}_{\sigma}({\bf k}) +
 \hat\Phi_{\sigma}({\bf k},\omega) /\hat{Q} .
 \label{27}
\end{eqnarray}
\par
The Dyson equation  (\ref{14})  with the zero-order quasiparticle
(QP) excitation energy (\ref{12}) and the self-energy (\ref{23})
gives an exact representation for the GF (\ref{9}).  To obtain a
closed system of equations, the multiparticle GF in the
self-energy operator (\ref{15}) should be evaluated as discussed
below.

\section{Approximate system of equations}
\label{sec:4}

In this section we derive an approximate system of equations for
the GFs  and the self-energy components in Eqs.~(\ref{24}) --
(\ref{27}) for the two Hubbard subbands adopting several
approximations to make the system of equations numerically
tractable.

\subsection{Generalized mean-field approximation}
\label{sec:4a}

The energy matrix (\ref{12}) is calculated using the commutation
relations (\ref{6}) for the HOs. The normal part of the energy
matrix $\hat{\varepsilon}({\bf k}) $ after diagonalization
determines the QP spectrum in  two Hubbard subbands in the GMFA
(for detail see~\cite{Plakida07}):
\begin{eqnarray}
{\varepsilon}_{1, 2} ({\bf k})& = & ({1}/{2}) [\omega_{2} ({\bf
k}) + \omega_1 ({\bf k})] \mp({1}/{2}) \Lambda({\bf k}),
 \label{17}\\
 {\omega}_\iota({\bf k})& = &
 4  t\,\alpha_{\iota} \gamma({\bf k})
 + 4 \,\beta_{\iota}\,t'\gamma'({\bf k})+
 4 \,\beta_{\iota}\,t''\gamma''({\bf k})
 \nonumber   \\
& + &\omega^{(c)}_\iota({\bf k}) + U \delta_{\iota,2} - \mu,
 \quad (\iota = 1, 2)
\label{17a}\\
  \Lambda({\bf k}) &= &    \{[\omega_{2} ({\bf k})
  - \omega_1 ({\bf k})]^2 + 4 W({\bf k})^2 \}^{1/2},
\nonumber\\
 W({\bf k}) & = &  4  t\,\alpha_{12} \gamma({\bf k})
 + 4 t' \,\beta_{12} \gamma'({\bf k})+
 4 t'' \,\beta_{12} \gamma''({\bf k}).
\nonumber
\end{eqnarray}
Here  the hopping parameter is defined by the expression:
\begin{eqnarray}
t_{ij} & = & (1/N)\,\sum_{\bf k}\exp[i{\bf k (i-j)}]\, t({\bf k}),
\label{17b}\\
t({\bf k})& =&  4 t \, \gamma({\bf k}) + 4 t' \,\gamma'({\bf k})
+ 4 t''\, \gamma''({\bf k}),
 \label{17c}
\end{eqnarray}
where the nearest-neighbor  hopping is  $t\,$, diagonal hopping
is $\, t'\,$ and the third neighbor hopping is $ \, t''\,$. The
corresponding  ${\bf k}$-dependent functions are: $\,\gamma({\bf
k})= (1/2)(\cos k_x +\cos k_y), \; \gamma'({\bf k}) = \,\cos k_x
\cos k_y \, $, and $\; \gamma '' ({\bf k})= (1/2)(\cos 2 k_x
+\cos 2 k_y) $ (the lattice constants $ a_{x}= a_{y}$ are put to
unity). The contribution from the  CI $V_{i j}$ in (\ref{17a}) is
given by
\begin{equation}
\omega^{(c)}_{1(2)}({\bf k})= \frac{1}{N } \sum_{\bf q} V({\bf k
-q}) N_{1(2)}({\bf q}),
 \label{17ci}
 \end{equation}
where $ N_1({\bf q}) = \langle X_{\bf q}^{0 \bar\sigma}X_{\bf
q}^{\bar\sigma 0}\rangle / Q_1\,$ and $\,N_2({\bf q})=
 \langle X_{\bf q}^{\sigma 2}X_{_{\bf q}}^{2\sigma}\rangle / Q_2\,$
are  occupation numbers in the single-particle and two-particle
subbands, respectively. $V({\bf q})$ is the Fourier transform of
$V_{ij}$.
\par
The kinematic interaction for the HOs results in renormalization
of  the spectrum (\ref{17}) determined by the parameters: $\,
\alpha_{\iota}= Q_{\iota}[ 1 + {C_{1}}/{Q^2_{\iota}}], \,
\beta_{\iota} = Q_{\iota}[ 1 + {C_{2}}/{Q^2_{\iota}}]\,$, $\,
\alpha_{12}= \sqrt{Q_{1}Q_{2}}[ 1 - {C_{1}}/{Q_{1}Q_{2}}] ,\,
\beta_{12} = \sqrt{Q_{1}Q_{2}}[ 1 -{C_{2}}/{Q_{1}Q_{2}}]\,$. In
addition to the conventional Hubbard I renormalization given by
$Q_{1}, \, Q_{2}$  parameters an essential   renormalization is
caused by the AF spin correlation functions  for
nearest-neighbors  and next neighbors, respectively:
\begin{equation}
C_{1} = \langle {\bf S}_i{\bf S}_{i+ a_1} \rangle, \quad C_{2} =
\langle {\bf S}_i{\bf S}_{i+ a_2}\rangle .
 \label{18}
\end{equation}
These functions strongly depend on doping resulting in a
considerable variation of the electronic spectrum  as shown later
and discussed in  detail  in Ref.~\cite{Plakida07}.
\par
The anomalous component $\, \hat{\Delta}_{\sigma}({\bf k}) \, $
of the matrix (\ref{12}) determines the superconduction gap in
the GMFA.   The diagonal matrix components in the coordinate
representation are given by the equations:
\begin{eqnarray}
&&   \Delta\sb{ij\sigma}\sp{22} Q_2 =  -  \sigma\, t_{ij}^{21}
   \langle X\sb{i}\sp{02} N\sb{j} \rangle  - V_{i j}
 \langle  X_{i}^{\sigma 2}\, X_{j}^{\bar\sigma 2} \rangle ,
   \label{19a}\\
&&  \Delta\sb{ij\sigma}\sp{11} Q_1 =   \sigma \, t_{ij}^{12}
   \langle N\sb{j}  X\sb{i}\sp{02} \rangle - V_{i j}
   \langle  X_{i}^{0 \bar\sigma} X_{j}^{0\sigma}  \rangle .
 \label{19b}
\end{eqnarray}
Here we introduced upper indexes for the hopping parameter
$\,t_{ij}^{12}, \, t_{ij}^{21}$ to stress that   the anomalous
components $\langle X\sb{i}\sp{02} N\sb{j} \rangle$ are induced
by the interband hopping.   Calculation of the correlation
function $\, \langle X\sb{i}\sp{02} N\sb{j} \rangle$ from the
equation of motion for the GF $\, L_{ij}(t-t') = \langle \langle
X_{i}^{02} (t) \mid N_j (t') \rangle \rangle \, $ results in the
superconducting gap in the two-particle subband (for detail see
Ref.~\cite{Plakida03}):
\begin{eqnarray}
\Delta^{22}_{ij\sigma}  =
 (J_{i j} - V_{i j})\,\langle X_{i}^{\sigma2}
 X_{j}^{\bar\sigma2}\rangle /Q_2 ,
 \label{21}
\end{eqnarray}
where $\, J_{i j} = {4\, ( t_{ij}^{12})^2}/{U }$ is the AF
exchange interaction.  A similar equation holds for the gap in
the single-particle subband:
 $\, \Delta_{ij\sigma}^{11}= (J_{i j} -
 V_{i j})\,\langle X_{i}^{0\bar\sigma}
 X_{j}^{0\sigma} \rangle/Q_1 \,$.
Therefore, the pairing in the Hubbard model in the GMFA  is
similar to  the superconductivity in the $t$--$J$ model mediated
by the AF exchange interaction $J_{i j}$.

\subsection{Self-energy operator}
\label{sec:5}

The self-energy matrix~(\ref{23}) due to the kinematic
interaction, as shown in Eq.~(\ref{7}), is determined by
multiparticle GFs such as $\, \langle \!\langle \hat
X_{l\sigma'}(t)\, B\sb{i\sigma\sigma'}(t)\, \mid
 \hat X_{l'\sigma''}^{\dagger}\,
B\sb{j\sigma\sigma''}^\dag \rangle \!\rangle \,$. We  calculate
the self-energy matrix in the SCBA using the mode-coupling
approximation  for the multiparticle GFs. In this approximation,
a propagation of excitations described by the Fermi-like
operators $\,\hat X_{l\sigma} \,$ and  the Bose-like operators
$B\sb{i\sigma\sigma'}$ for $l \neq i$ is assumed to be
independent.  Therefore, the corresponding time-dependent
multiparticle correlation functions can be written   as a product
of fermionic and bosonic correlation functions,
\begin{eqnarray}
 &&\langle X_{l'}^{2\sigma''} B\sb{j\sigma\sigma''}^\dag
 |B\sb{i\sigma\sigma'}(t)
X_l^{\sigma' 2}(t)\rangle
\nonumber \\
 && = \delta_{\sigma', \sigma''}\langle X_{l'}^{2\sigma'}
X_l^{\sigma' 2}(t)\rangle \langle B\sb{j\sigma\sigma'}^\dag
 |B\sb{i\sigma\sigma'}(t) \rangle,
\label{B5} \\
&& \langle  X_{l'}^{\bar{\sigma}'' 2} B\sb{j
\bar{\sigma}\bar{\sigma}''}
 | B\sb{i\sigma \sigma'}(t) X_l^{\sigma' 2}(t)\rangle
\nonumber \\
 && = \delta_{\sigma', \sigma''}
 \langle  X_{l'}^{\bar{\sigma}' 2}
  X_l^{\sigma' 2}(t)\rangle\,
  \langle   B\sb{j\bar{\sigma}\bar{\sigma}'}
  B\sb{i\sigma \sigma'}(t) \rangle \, .
 \label{B6}
\end{eqnarray}
The time-dependent single-particle correlation functions are
calculated self-consistently using the corresponding GFs. This
approximation results in a self-consistent system of equations
for the self-energy  (\ref{23}) and the  GFs (\ref{24}),
(\ref{25}) similar to the strong-coupling  Eliashberg
theory~\cite{Eliashberg60} (for detail see Ref.~\cite{Plakida13}
and Chapter~A in Ref.~\cite{Plakida10}).
\par
In this approximation  the normal state  GF (\ref{26}) for two
subbands takes the form~\cite{Plakida07}:
\begin{eqnarray}
 {G}^{11(22)}_{N}({\bf k},\omega)& =&
[1 - b({\bf k})] {G}_{1(2)}({\bf k},\omega) \nonumber \\
 &+&   b({\bf k}){G}_{2(1)}({\bf k},\omega)  ,
 \label{33}\\
{G}_{1 (2)}({\bf k},\omega)& = & \frac{1}
 {\omega - {\varepsilon}_{1(2)}({\bf k}) -
 \Sigma({\bf k},\omega)} \, ,
 \label{34}
\end{eqnarray}
where the hybridization parameter $\,  b({\bf k}) =
[{\varepsilon}_{2} ({\bf k}) - \omega_{2}({\bf k})] /
[{\varepsilon}_{2} ({\bf k}) - {\varepsilon}_{1} ({\bf k})]\, $.
The self-energy $\Sigma({\bf k},\omega)$  can be approximated by
the same  function for two subbands.  In  the imaginary frequency
representation, $ i\omega_{n}=i\pi T(2n+1)$, $n = 0,\pm 1, \pm 2,
...\, $ it reads
\begin{eqnarray}
{\Sigma}({\bf k}, \omega_{n}) & = & - \frac{T}{N}\sum_{\bf q}
 \sum_{m}\lambda^{(+)}({\bf q, k-q} \mid
\omega_{n}-\omega_{m})
 \nonumber \\
&\times &    [{G}_{1}({\bf q}, \omega_{m})+
  {G}_2({\bf q}, \omega_{m})]
\nonumber\\
 &\equiv & i\omega_{n}\,[1-Z({\bf k},\omega_n)]
  + X({\bf k},\omega_n).
 \label{45a}
\end{eqnarray}
The normal GF (\ref{34}) for the two subbands  takes the form:
\begin{eqnarray}
\{G_{1(2)}({\bf k},  \omega_{n})\}^{-1}= i\omega_n  -
{\varepsilon}_{1(2)}({\bf k}) - \Sigma({\bf k},\omega_n)
 \nonumber \\
 = i\omega_n Z({\bf k},\omega_n) -[{\varepsilon}_{1(2)}({\bf k})+
X({\bf k},\omega_n)] \, .
 \label{45}
\end{eqnarray}
To calculate $T_c$ we can use a linear approximation for the pair
GF (\ref{25}). In particular, Eq.~(\ref{27})  for the
two-particle subband gap $\varphi({\bf k},\omega)= \sigma
\varphi_{2, \sigma}({\bf k},\omega)$ can be written as
\begin{eqnarray}
\varphi({\bf k}, \omega_n) & = &
  \frac{T_c}{N}\sum_{\bf q} \,  \sum_{m}\,
\{\, J({\bf k-q}) - V({\bf k-q})
\nonumber \\
& +  & \lambda^{(-)}({\bf q, k-q} \mid \omega_{n}-\omega_{m}) \}
 \label{46} \\
 &\times & \frac{[1 - b({\bf q})]^2\,
  \varphi({\bf q}, \omega_{m})}{[\omega_m Z({\bf q},\omega_m)]^2
  + [{\varepsilon}_{2}({\bf q})+
X{\bf q},\omega_m)]^2}\; .
 \nonumber
 \end{eqnarray}
The  interaction functions in (\ref{45a}) and (\ref{46}) in the
imaginary frequency representation are given by
\begin{eqnarray}
 && \lambda^{(\pm)}({\bf q },{\bf k -q } | \nu_n)  = -
|t({\bf q})|^{2} \, \chi\sb{sf}({\bf k- q},\nu_n)
\nonumber \\
&& \mp \{ |V({\bf k -q})|^2 + | t({\bf q})|^{2}/4 \}\,
\chi_{cf}({\bf k-q}, \nu_n) .
  \label{47}
\end{eqnarray}
The  spectral densities of bosonic excitations are determined by
the dynamic susceptibility for spin $( sf)$ and number (charge)
$( cf)$  fluctuations
\begin{eqnarray}
\chi\sb{sf}({\bf q},\omega) & = &
  -  \langle\!\langle {\bf S\sb{q} | S\sb{-q}}
\rangle\!\rangle\sb{\omega},
\label{32a} \\
 \chi\sb{cf}({\bf q},\omega) &= &
 - \langle\!\langle \delta N\sb{\bf q} | \delta N\sb{-\bf q}
   \rangle\!\rangle\sb{\omega} ,
\label{32b}
\end{eqnarray}
written in terms of the commutator GFs~\cite{Zubarev60} for  spin
${\bf S \sb{q}} $ and number $\delta N_{\bf q} = N_{\bf q} -
\langle N_{\bf q} \rangle$ operators.
\par
Thus, we have derived  the self-consistent system of equations
for the normal GF (\ref{45}), the self-energy (\ref{45a}), and
the gap function (\ref{46}). In the SCBA, vertex corrections to
the kinematic  interaction $\, t({\bf q}) \,$ of electrons with
spin- and charge-fluctuations (\ref{32a}), (\ref{32b}) induced by
the intraband hopping are neglected. It is assumed that the
system is far away from a charge instability or a stripe
formation and  charge-fluctuations give a small contribution to
the pairing. The largest contribution from spin fluctuations
comes from  wave-vectors close the AF wave-vector  ${\bf Q} =
(\pi, \pi)$ where their  energy $\omega_{s}({ \bf Q})$  is much
smaller than the Fermi energy,  $\, \omega_{s}({ \bf Q}) /\mu \ll
1\,$  (see, e.g.,~\cite{Vladimirov09}). Therefore, vertex
corrections to the kinematic interaction should be small as in
Eliashberg theory~\cite{Eliashberg60} for electron interaction
with phonons, where $\, \omega_{ph}({\bf q})/\mu \ll 1\,$.
Consequently,  the SCBA for the self-energy and the GFs
calculated self-consistently is quite reliable and makes it
possible to consider the strong coupling regime  which is
essential in study of renormalization of the QP spectrum and  the
superconducting pairing as shown in
Refs.~\cite{Plakida07,Plakida13} and discussed later.

\section{Results and discussion}
\label{sec:7}

In numerical computations  we have used   models  for the CIs and
the susceptibility (\ref{32a}), (\ref{32b}).  For the intersite
CI $\,V_{ij} \,$ we consider a model for repulsion of two
electrons (holes) on neighbor lattice sites,
\begin{equation}
V({\bf q}) = 2 V\, (\cos q_x + \cos q_y )\, ,
 \label{50}
\end{equation}
with various values of  $\,  V = 0.0,\, 0.5\,t,\, 1.0\,t\,$ and
$  2.0 \,t\,$.  For the on-site CI we consider  $U = 8\,t, \,16
\,t$ and $ \, 32\,t \,$. The AF exchange interaction for neighbor
sites is described by the function $\,J({\bf q})= 2 J\, (\cos q_x
+ \cos q_y ) $. Note, that in the GMFA  the CI $V_{ij}$ gives no
contribution to the exchange interaction $\, J_{i j}$ and
therefore  it is assumed to be the same for all values of $V\,$
(cf. with Refs.~\cite{Plekhanov03,Senechal12}). In the most of
calculations we take $\,J = 0.4 t\,$ but to study a role  of the
spin-fluctuation interaction in the superconducting pairing, we
consider also other values of the interaction, $\,J = 0.2 \,t,\,
0.6\, t,\,$ and  $1.0 \,t\,$.
\par
Due to a large energy scale of charge fluctuations, of the order
of several $\,t$,  in comparison with the spin excitation energy
of the order of $J$, the charge fluctuation contributions  can be
considered in the static limit for the susceptibility (\ref{32b})
\begin{eqnarray}
\chi_{cf}({\bf k}) &=& \chi_{cf}^{(1)}({\bf k})
  + \chi_{cf}^{(2)}({\bf k}), \label{59} \\
\chi_{cf}^{(\alpha)}({\bf k}) &=& - \frac{1}{N}\sum_{\bf q}
 \frac{N^{(\alpha)}({\bf q +k}) - N^{(\alpha)}({\bf q})}
 {\varepsilon_{\alpha}({\bf q +k}) -
 \varepsilon_{\alpha}({\bf q})},
 \nonumber
\end{eqnarray}
where the   occupation numbers $ N^{(\alpha)}({\bf q})$ are
defined as
\begin{eqnarray}
N^{(1)}({\bf k}) &=& [Q_1 + (n-1)b({\bf k})]\, {N}_{1}({\bf k}),
\nonumber \\
N^{(2)}({\bf k})  &=& [ Q_2 - (n-1)b({\bf k})]\, {N}_{2}({\bf k}),
\nonumber \\
{N}_{\alpha}({\bf k})&=&  ({1}/{2}) + T\sum_{m} \,
   G_{\alpha}({\bf k}, \omega_{m}).
  \label{38}
\end{eqnarray}
\par
For the dynamical spin susceptibility $\, \chi_{sf}({\bf
q},\omega)$ (\ref{32a}) we used a model suggested in
Ref.~\cite{Jaklic95}
\begin{eqnarray}
  && {\rm Im}\, \chi_{sf}({\bf q},\omega+i0^+) =
 \chi_{sf}({\bf q}) \; \chi_{sf}''(\omega)
 \nonumber \\
& = &\frac {\chi_{ Q}}{1+ \xi^2 [1+ \gamma({\bf q})]} \;  \tanh
\frac{\omega}{2T} \frac{1}{1+(\omega/\omega_{s})^2}\, .
 \label{51}
\end{eqnarray}
This type of the spin-excitation spectrum was found in the
microscopic theory for the $t$-$J$ model in
Ref.~\cite{Vladimirov09}. The model is determined by two
parameters: the AF correlation length $\xi$ and  the cut-off
energy of spin excitations of the order of the exchange energy
$\omega_s \sim J$. The strength of the spin-fluctuation
interaction given by the static susceptibility $\chi_{ Q} =
\chi_{sf}({\bf Q}) $ at the AF wave vector ${\bf Q = (\pi,\pi)}$,
\begin{equation}
\chi_{ Q} = \frac{3 (1- \delta)}{2\omega_{s} }
 \left\{ \frac{1}{N} \sum_{\bf q}
 \frac{1}{ 1+\xi^2[1+\gamma({\bf q})]} \right\}^{-1} ,
 \label{52}
 \end{equation}
is defined by the normalization condition:
\begin{eqnarray}
 \frac{1}{N} \sum_{\bf q} \int\limits_{0}^{\infty}
  \frac{d \omega}{\pi}  \coth\frac{\omega}{2T} \, {\rm Im}
  \chi_{sf}({\bf q},\omega)=\langle {\bf S}_{i}^2\rangle
  = \frac{3}{4}(1- \delta). &&
  \nonumber
\end{eqnarray}
The spin correlation functions (\ref{18}) in the single-particle
excitation spectrum (\ref{17}) are calculated using the same
model (\ref{51}): $\,
 C_1 = ({1}/{N}) \sum_{\bf q}\, C_{\bf q}\, \gamma({\bf q}),
 \;
 C_2 = ({1}/{N}) \sum_{\bf q} \, C_{\bf q}\, \gamma'({\bf q})
 \,$,
where  $\, C_{\bf q}= (\omega_{s}/ 2)
 (\chi_{ Q}\, / \{1+\xi^2[1+ \gamma({\bf q})]\}) $.
As an energy unit we use $t = 0.4$~eV and for the hopping
parameters we take  $t' = - 0.2 \, t, \quad t'' = 0.10 \, t$.
Below we present numerical  results for a hole-doped case for the
two-hole subband.

\subsection{Electronic spectrum in the normal state}
\label{sec:8}

\begin{figure}
\resizebox{0.35\textwidth}{!}{%
\includegraphics{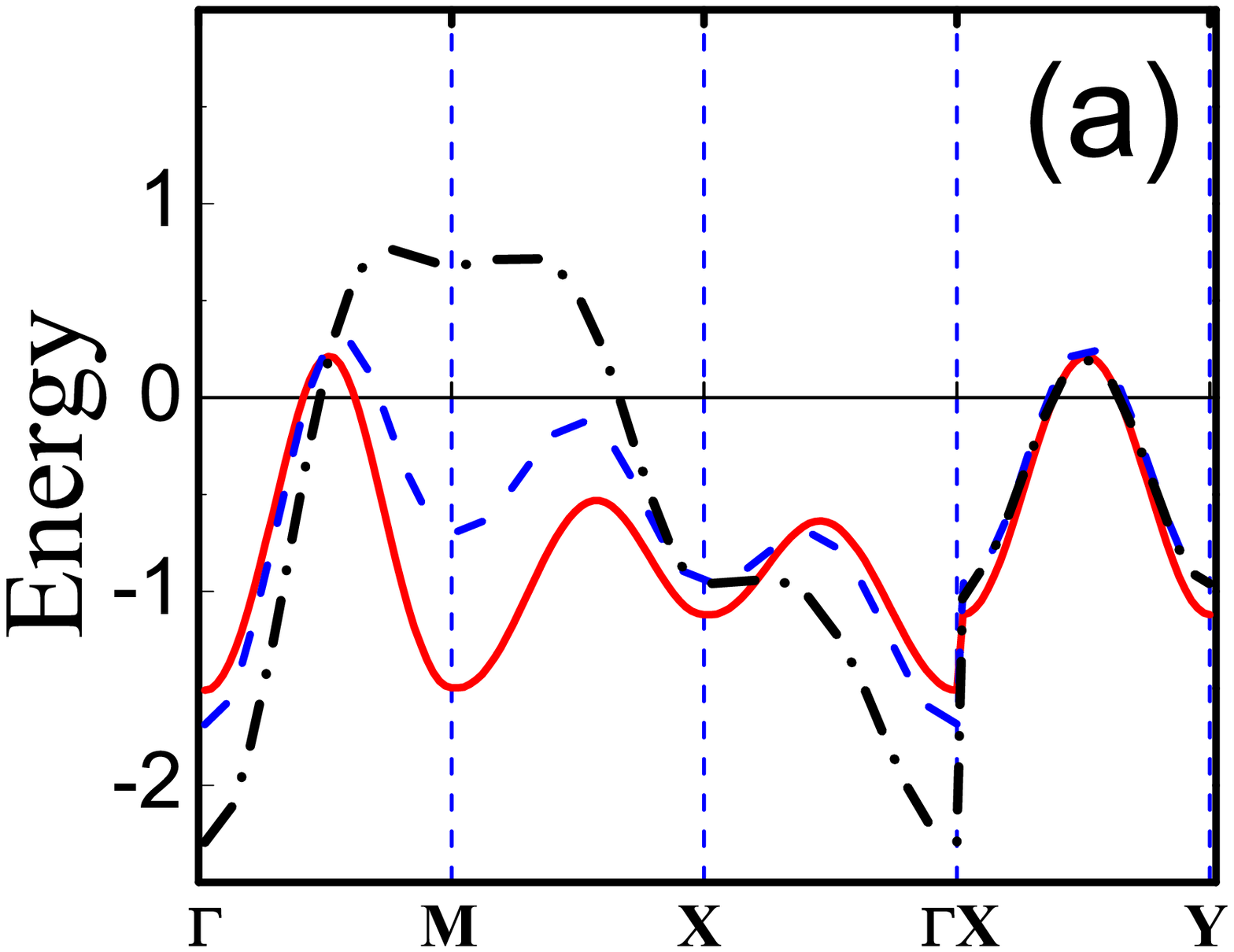}}\vspace{5mm}
\resizebox{0.35\textwidth}{!}{%
\includegraphics{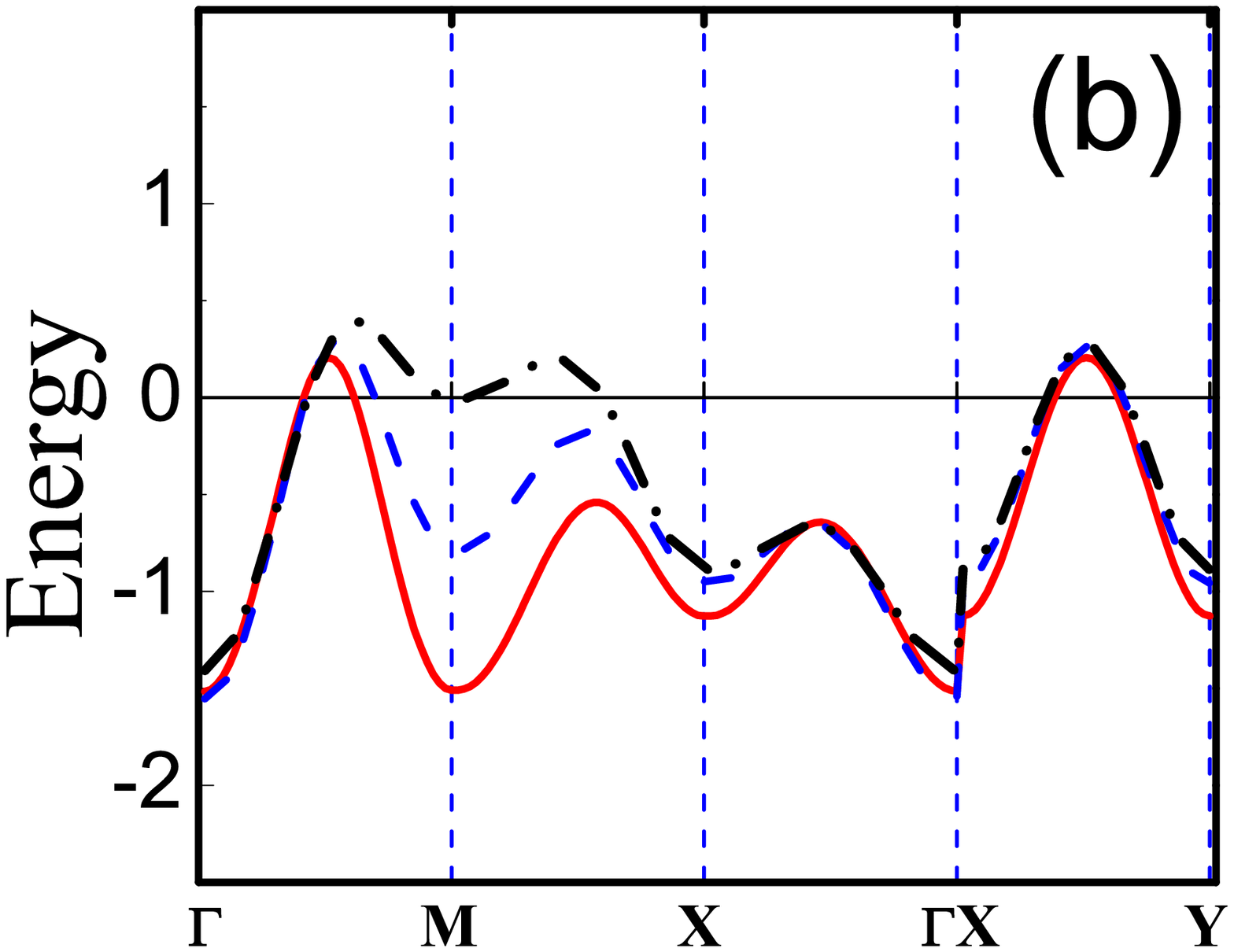}}
 \caption{(Color online) Electron dispersion in the GMFA
${\varepsilon}_{2} ({\bf k})$ for (a) $V=0$ and (b)  $V=2$ at $U
= 8$  along the symmetry directions $\Gamma(0, 0)\rightarrow
M(\pi,\pi) \rightarrow X (\pi, 0) \rightarrow \Gamma(0, 0)$ and
$X (\pi, 0) \rightarrow Y(0, \pi)$ for $\delta = 0.05\,$ (red
solid line), $0.10$ (blue dashed  line), and $0.25$ (black
dash-dotted line). Fermi energy for hole doping is at $\omega =
0$.}
 \label{Ek8}
\end{figure}

\begin{figure}
\resizebox{0.35\textwidth}{!}{%
\includegraphics{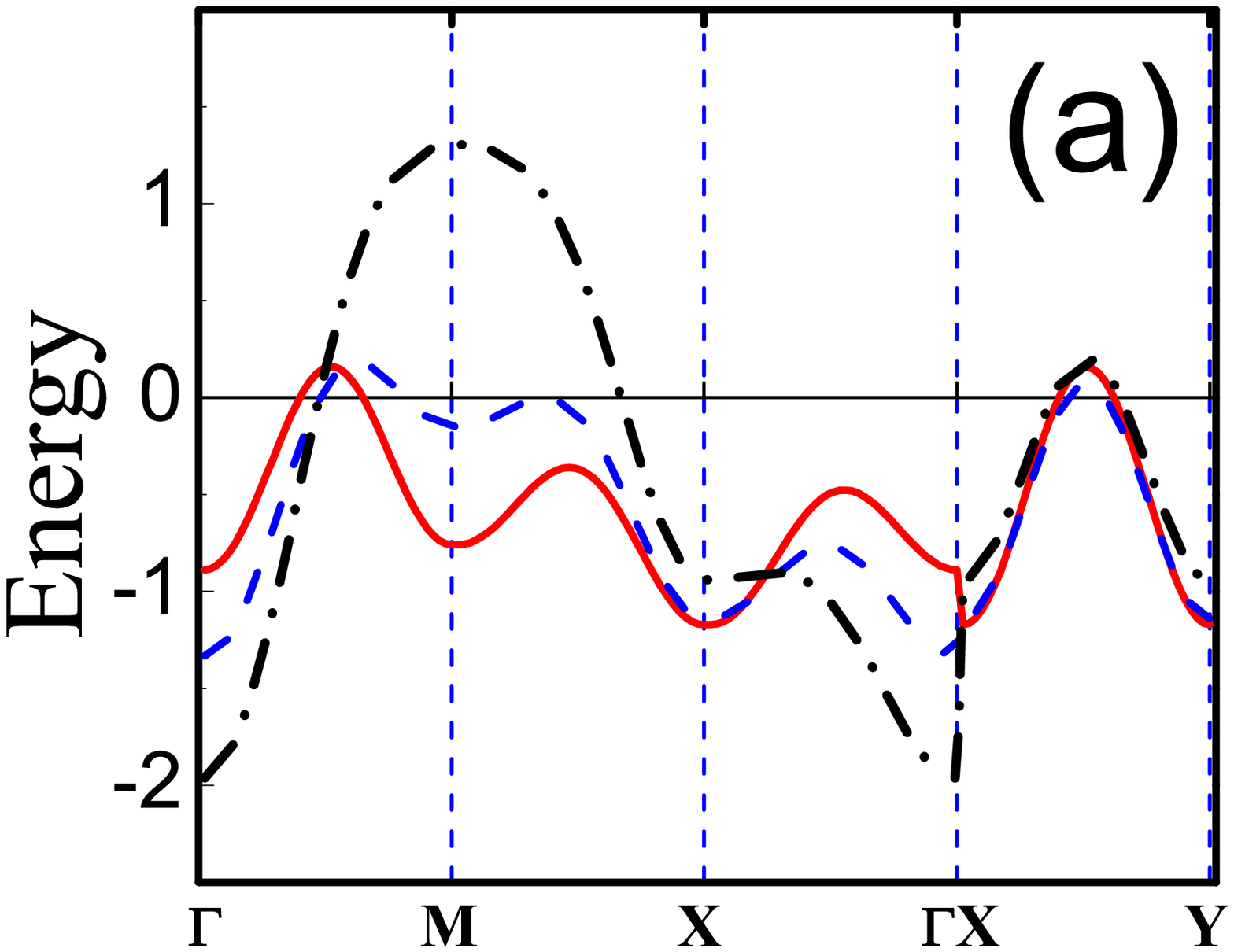}}\vspace{5mm}
\resizebox{0.35\textwidth}{!}{%
\includegraphics{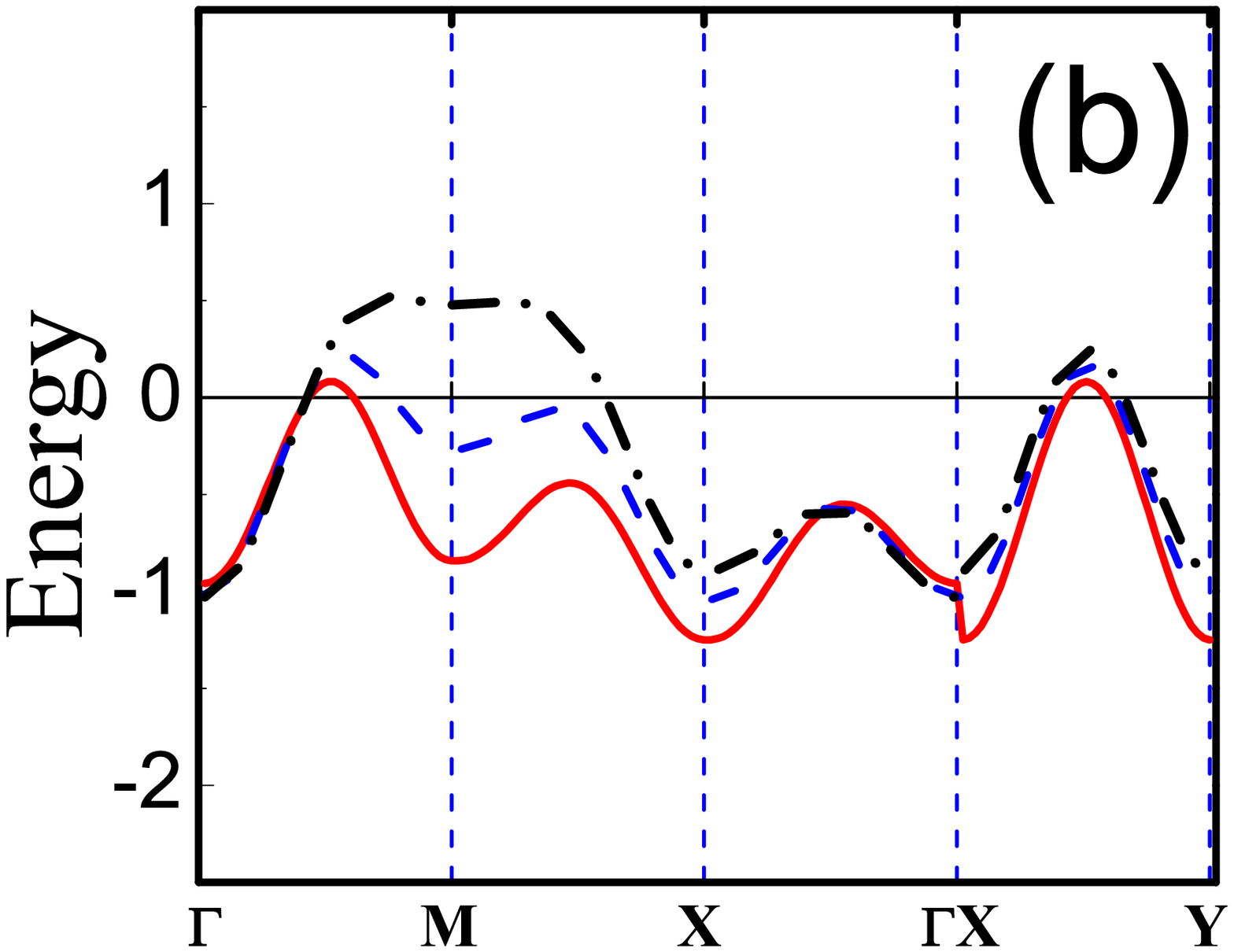}}
 \caption{(Color online) The same as in Figure~\ref{Ek8} for  $U = 16$.}
 \label{Ek16}
\end{figure}

\begin{figure}[ht!]
\resizebox{0.35\textwidth}{!}{%
\includegraphics{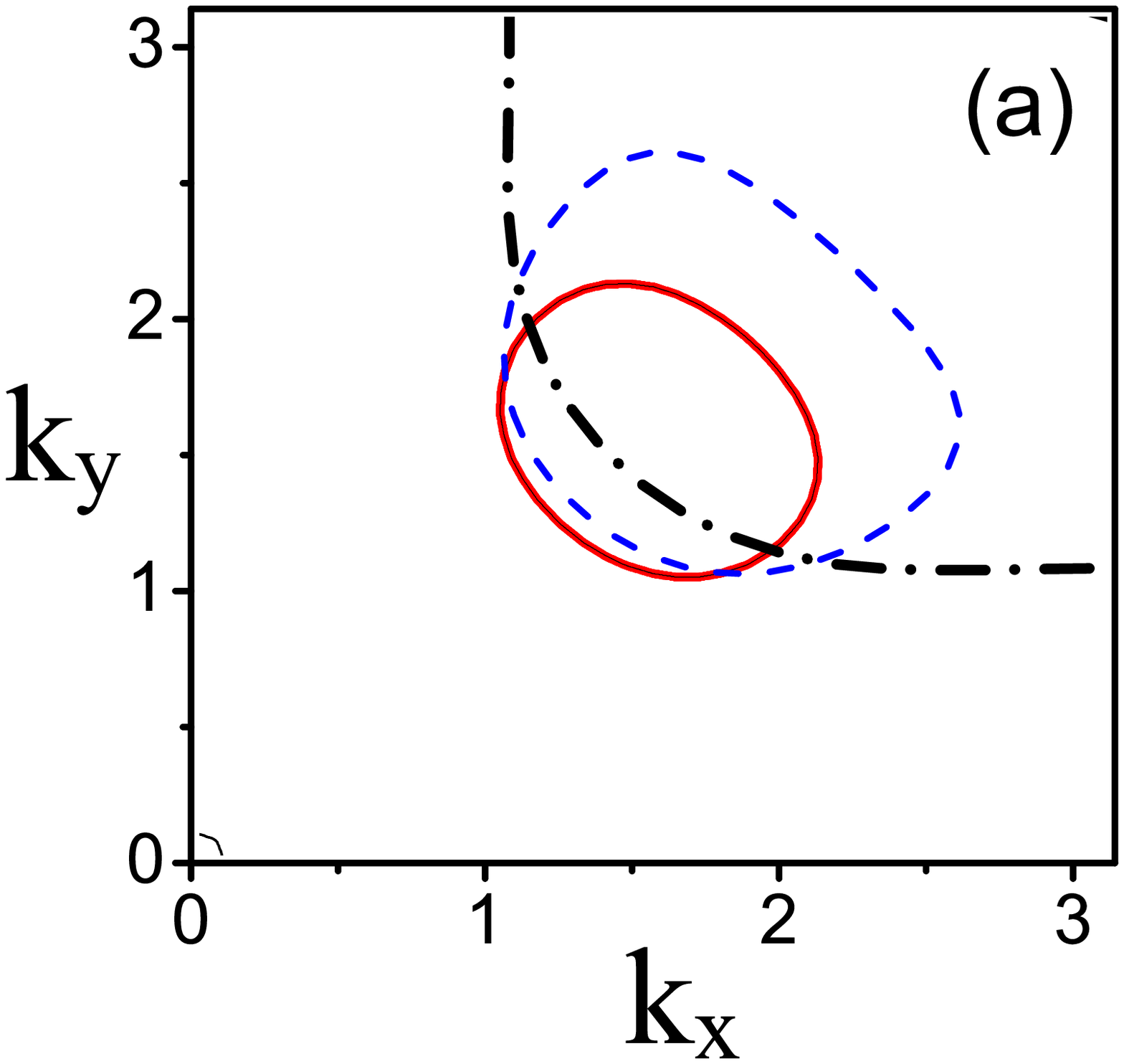}}\vspace{5mm}
\resizebox{0.35\textwidth}{!}{%
\includegraphics{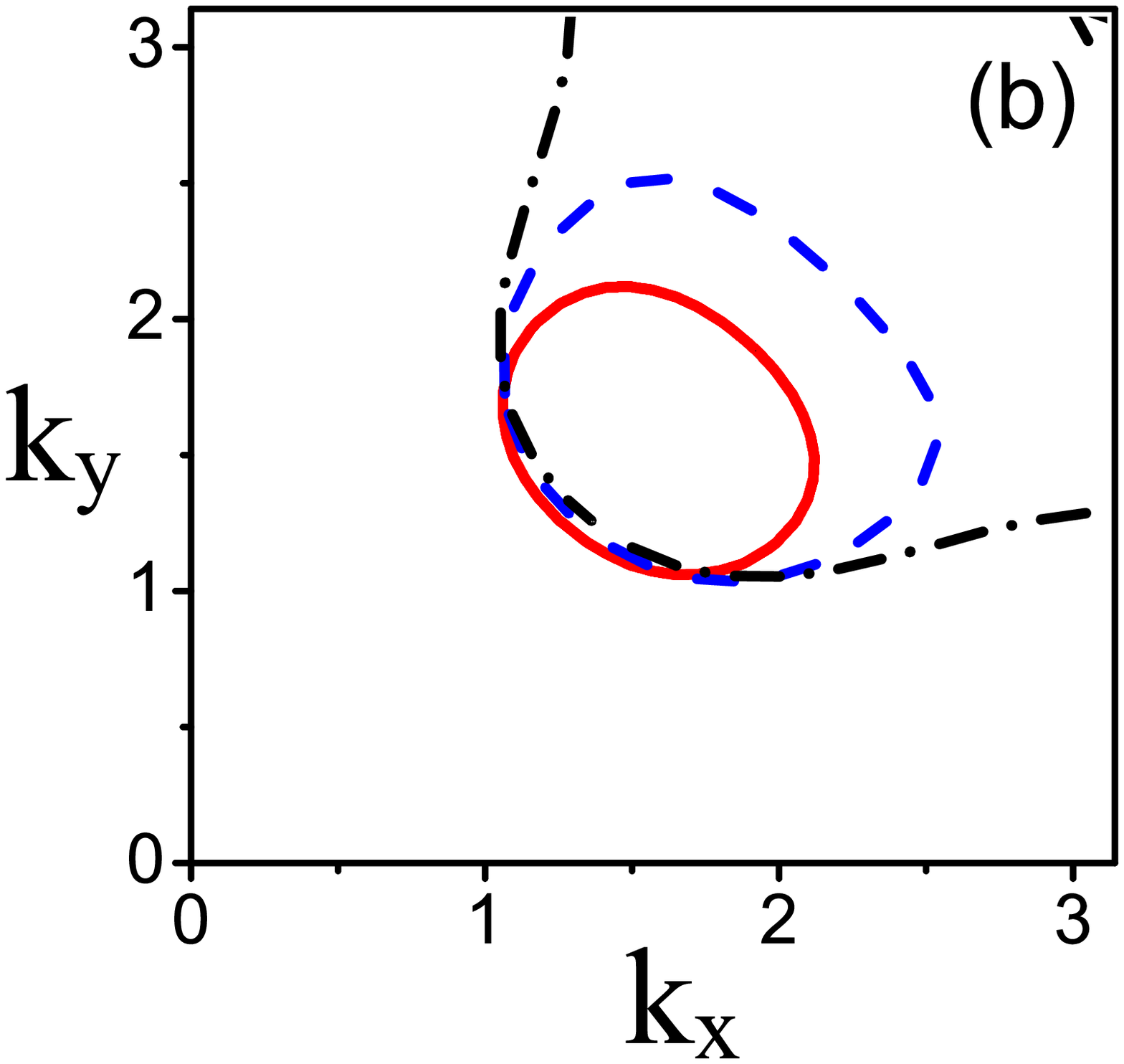}}
\caption{(Color online)  Fermi surface for (a) $V=0$ and (b)
$V=2$ at $U = 8$ in  the quarter of the BZ in the GMFA at hole
doping $\delta = 0.05$ (red solid line), $ 0.10$ (blue dashed
line), and $ 0.25$ (black dash-dotted line).}
 \label{FS8}.
\end{figure}

\begin{figure}
\resizebox{0.35\textwidth}{!}{%
\includegraphics{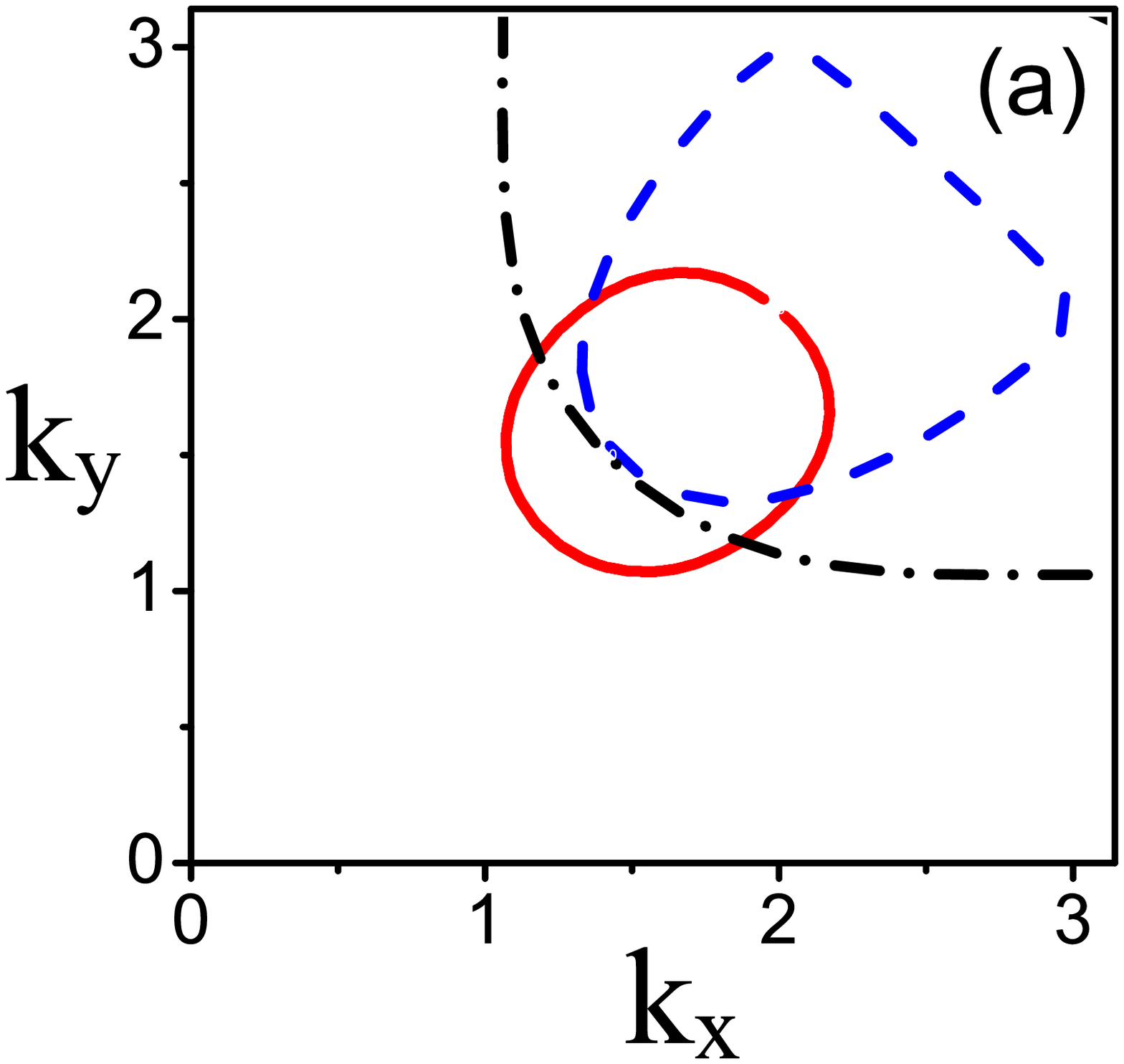}}\vspace{5mm}
\resizebox{0.35\textwidth}{!}{%
\includegraphics{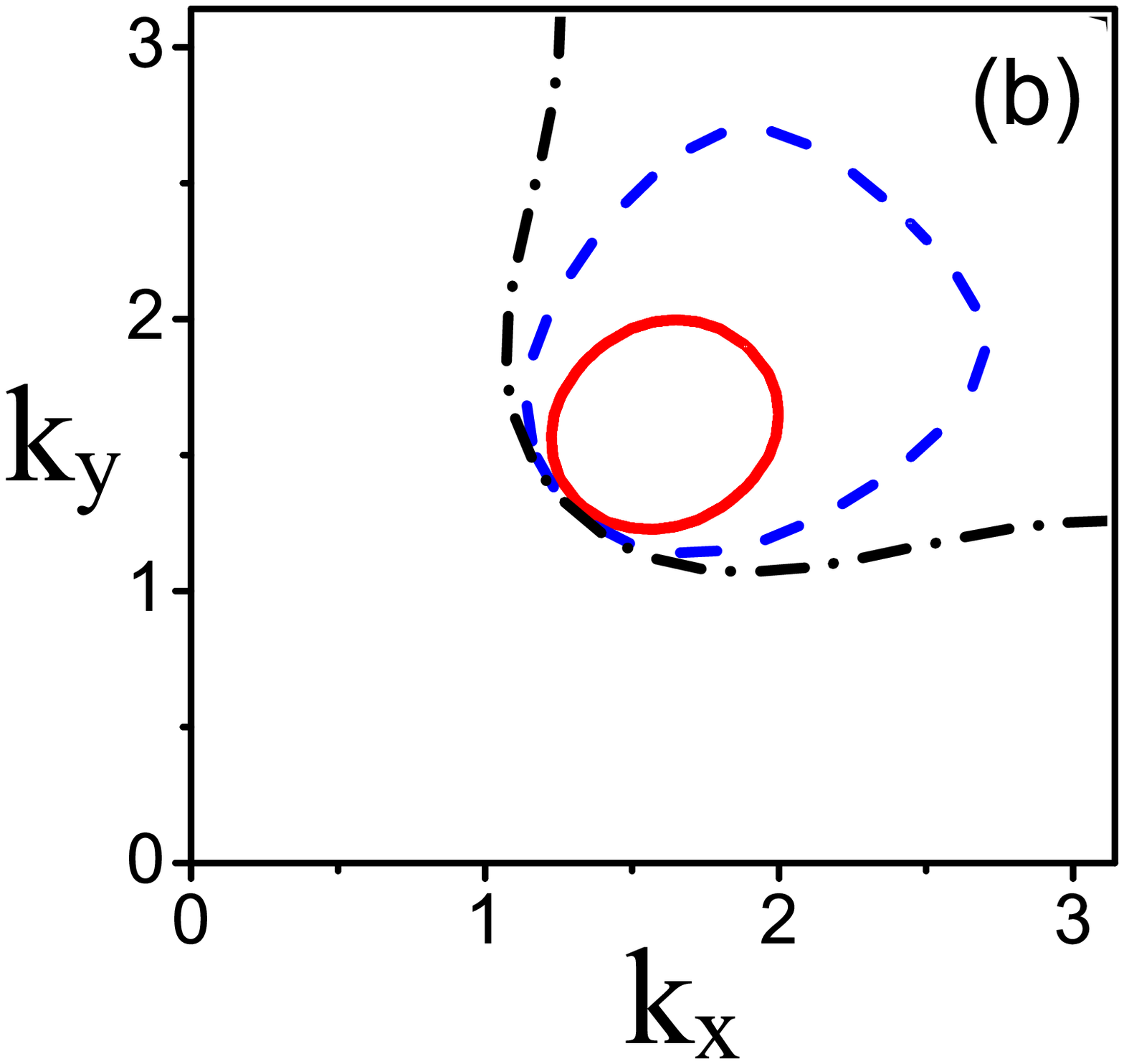}}
  \caption{(Color online)  The same as in Figure~\ref{FS8} for  $U = 16$.}
 \label{FS16}.
\end{figure}

At first we consider  results in the GMFA for the electronic
spectrum (\ref{17}). The doping dependence of the electron
dispersion for the two-hole subband  ${\varepsilon}_{2} ({\bf
k})$  along the symmetry directions in the 2D Brillouin zone (BZ)
are shown in Fig.~\ref{Ek8}  for $U = 8$ and in  Fig.~\ref{Ek16}
for $U = 16$    for $V = 0$ (a) and for $V = 2$ (b). The
corresponding  Fermi surfaces (FSs) determined by the equation:
$\, \varepsilon_{2}({\bf k_{\rm F}})   = 0 \,$ are  plotted in
Fig.~\ref{FS8} and Fig.~\ref{FS16}. For small doping, $\, \delta
= 0.05$, the energy at the $M(\pi,\pi)$ and $\Gamma(0, 0)$ points
are nearly equal as in the AF phase. Only small hole-like FS
pockets close to the $(\pm \pi/2,\pm \pi/2)$ points emerge at
this doping as shown in Figs.~\ref{FS8}, \ref{FS16}. With
increasing doping, the AF correlation length decreases that
results in increasing of the electron energy at the $M(\pi,\pi)$
point and at some critical doping $\delta \sim 0.12$ a large FS
appears. At the same time, the renormalized two-hole subband
width increases with doping, as e.g. for $U=8$ and $V = 0$ from
$\widetilde{W} \approx 2\,t$ at $\, \delta = 0.05\,$ to
$\widetilde{W} \approx 3\,t$ at $\, \delta = 0.25$, which,
however, remains less than the ``bare'' Hubbard subband width $W
= 4 t\,(1+\delta)$ where short-range AF correlations are
disregarded. With increasing CI $U$ and $V$ the subband width
shrinks as seen from comparison panels (a) and (b) for electronic
spectra  in Figs.~\ref{Ek8}, \ref{Ek16} and the FS  in
Figs.~\ref{FS8}, \ref{FS16}.

\begin{figure}
\resizebox{0.35\textwidth}{!}{%
\includegraphics{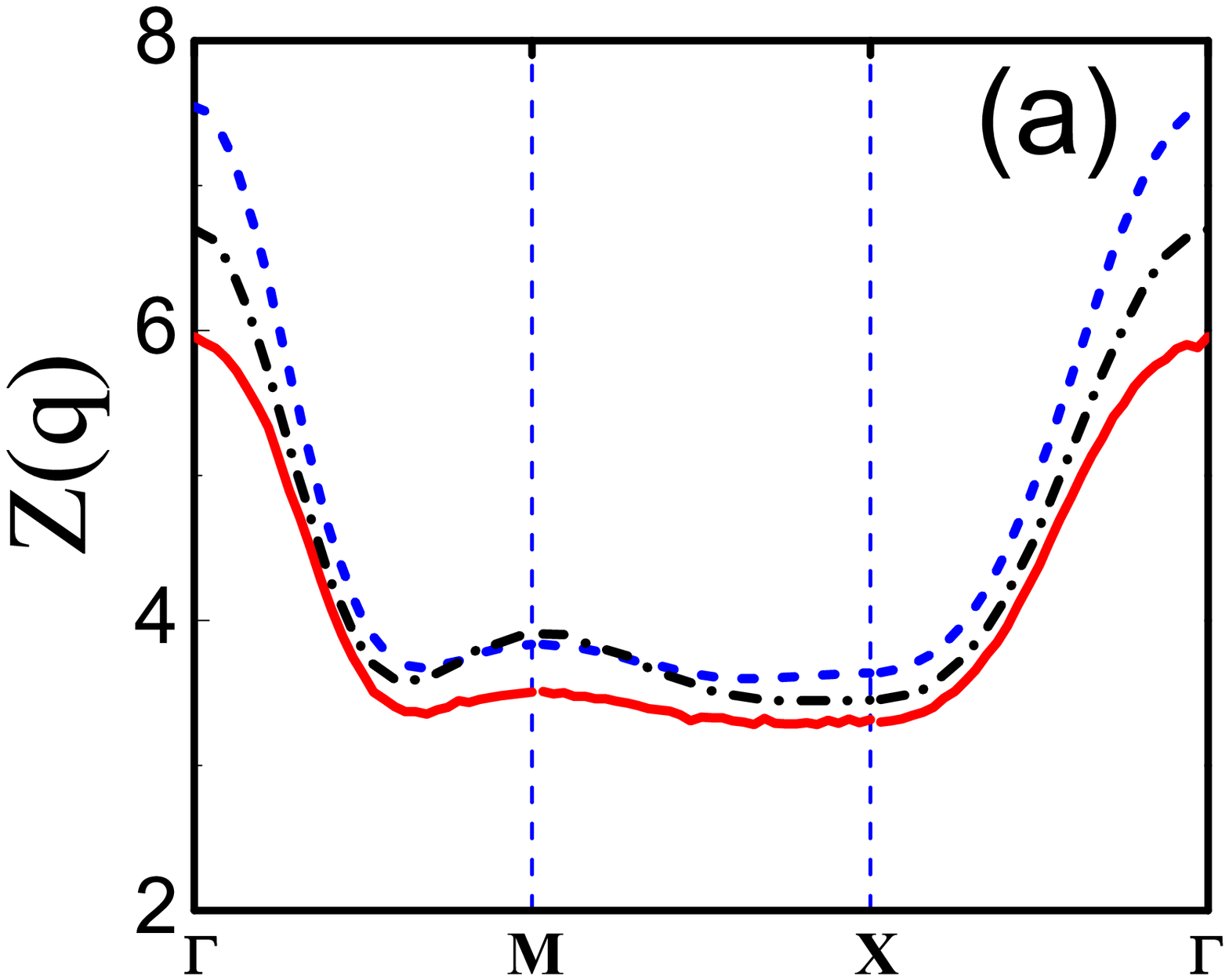}}
\resizebox{0.35\textwidth}{!}{%
\includegraphics{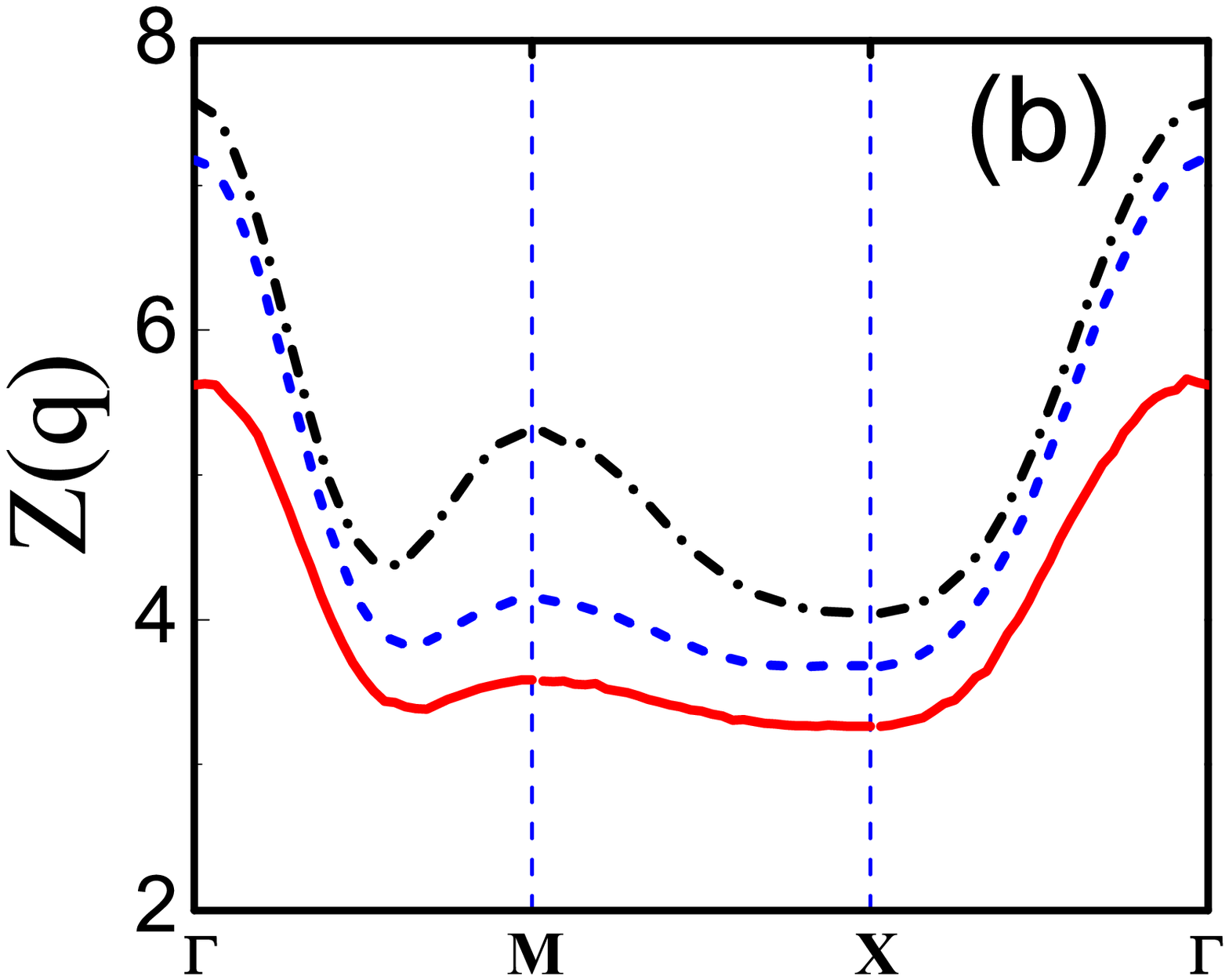}}
\caption{(Color online) The renormalization parameter $Z({\bf q})
$ along the symmetry directions $\Gamma(0, 0)\rightarrow
M(\pi,\pi) \rightarrow X (\pi, 0) \rightarrow \Gamma(0, 0)$ at
$\delta = 0.10$  at $U = 8$ (red solid line), $U = 16$ (blue
dashed line), and $U = 32$ (black dash-dotted line)  for (a)
$V=0$ and (b)  $V=1$.}
 \label{Zq}
\end{figure}

To study  self-energy effects in  the electronic spectrum the
strong coupling theory (SCT) should be used as  a self-consistent
solution of the system of equations for the normal GF (\ref{34})
and  the self-energy (\ref{45a}). Since  detailed investigation of
the normal state electronic spectrum in SCT was performed for the
conventional Hubbard model in Ref.~\cite{Plakida07}  and for the
extended Hubbard model in Ref.~\cite{Plakida13},  here we present
results only for $U$ and $V$ dependence of the renormalization
parameter $Z({\bf q})$  at the Fermi energy
 \begin{eqnarray}
 Z({\bf q})& = &Z({\bf q}, \omega =0)= 1 + \lambda({\bf q})
 \nonumber \\
 & =&    1 - [\, d \,{\rm Re}\,
  \Sigma({\bf q}, \omega)/{d \omega}]|_{\omega = 0}  \,  .
 \label{45ba}
 \end{eqnarray}
We found  that $Z({\bf q}) $  weakly depends on $\delta$ for
$\delta \lesssim 0.15$ (see also Ref.~\cite{Plakida13}).
Therefore,  in Fig.~\ref{Zq}  we demonstrate   the $U$ dependence
of $Z({\bf q}) $ at $\delta = 0.10$ for $V=0$~(a) and for
 $V=1$~(b).  It appears that the  renormalization parameter $Z({\bf q}) $ is
quite large in the whole BZ , $\, Z({\bf q}) \sim 4 -6\,$, which
results in a strong suppression of the QP weight $\, \sim
1/Z({\bf q})$.

\subsection{Superconducting $\bf T_c$}
\label{sec:9}

For a comparison of various contributions  to the superconducting
gap equation (\ref{46}),  we approximate  the interaction
(\ref{47}) by   its value close to the Fermi energy. As the
result instead of the dynamical susceptibility (\ref{32a}),
(\ref{32b}) the static susceptibility  $\chi({\bf q}) ={\rm Re}
\, \chi({\bf q},\Omega = 0) $ appears in the gap equation. It
brings us to the BCS-type equation for the gap function
(\ref{46}) at the Fermi energy $\varphi({\bf k}) = \varphi({\bf
k},\omega=0)$:
\begin{eqnarray}
&&\varphi({\bf k}) =
  \frac{1}{N}\sum_{{\bf q}} \,
  \frac{[1 - b({\bf q})]^2\,
  \varphi({\bf q})}{[Z({\bf q})]^2 \; 2\widetilde{\varepsilon}({\bf q})}
    \tanh\frac{\widetilde{\varepsilon}({\bf q})}{2T_c}
 \big\{ J({\bf k-q})
\nonumber \\
&& - V({\bf k-q})+ \big[(1/4) |t({\bf q})|\sp{2} + |V({\bf k
-q})|^2\big] \chi\sb{cf}({\bf k -q})
 \nonumber \\
&& -  |t({\bf q})|\sp{2}\; \chi_{sf}({\bf k -q})\theta(\omega_{s}
-| \widetilde{\varepsilon}({\bf q})|) \big\}\,,
     \label{55}
\end{eqnarray}
where $\,  \widetilde{\varepsilon}({\bf q}) =
\varepsilon_{2}({\bf q})/Z({\bf q}) $ is the renormalized energy.
Whereas for the exchange interaction and CI there are no
retardation effects and the pairing occurs for  all electrons in
the two-particle subband, the spin-fluctuation contributions is
restricted  to the range of energies $\,\pm \omega_s$ near the
FS, as determined by the $\theta$-function.
\par
To estimate various contributions in the gap equation (\ref{55})
we consider a model $d$-wave gap function, $\varphi({\bf k}) =
(\Delta/2) \,\eta({\bf k})$ where $ \eta({\bf k}) = (\cos k_x -
\cos k_y)$. Then the  gap equation  can be written in the form
(for detail see Ref.~\cite{Plakida13}):
\begin{eqnarray}
&& 1 = \frac{1}{N } \sum_{\bf q}\frac{[1 - b({\bf q})]^2\,
  [\eta({\bf q})]^2}{[Z({\bf q})]^2 \; 2\widetilde{\varepsilon}({\bf q})}
    \tanh\frac{\widetilde{\varepsilon}({\bf q})}{2T_c}\big\{ J - V +  \widehat{V}\sb{cf}
 \nonumber \\
& &  + (1/4)\,|t({\bf q})|\sp{2} \widehat{\chi}\sb{cf}
   - |t({\bf q})|^{2}\, \widehat{\chi}\sb{sf}
   \theta(\omega_{s} -| \widetilde{\varepsilon}({\bf q})|) \big\}.
\label{56}
\end{eqnarray}
In this equation only $l=2$ components of the static
susceptibility  and CI give contributions
\begin{eqnarray}
 \widehat{V}\sb{cf} & = & \frac{1 }{N}\sum_{\bf k}|V({\bf k})|^2
 \, \chi\sb{cf}({\bf k})\, \cos k_x ,
\label{57b} \\
\widehat{\chi}\sb{cf} & = &\frac{1}{N}\sum_{\bf k}
 \chi\sb{cf}({\bf k}) \cos k_x ,
\label{57c} \\
   \widehat{\chi}\sb{sf} & = &  \frac{1}{N}\sum_{\bf k}
 \chi\sb{sf}({\bf k})\cos k_x \, .
\label{57e}
\end{eqnarray}
\begin{table}
\caption{ Charge-fluctuation contribution $\,\widehat{V}_{cf}\,/t
\quad \; $ for several values of the on-site CI $\; U \;$ and
intersite CI $\; V \; $ for  hole concentrations $\; \delta =
0.10$.}
 \label{Table1}
\begin{tabular}{crrrc}
\hline \hline $\; U \quad $  &   $\; V = 1  \quad $ &   $ \;
V = 2  \quad  $ &   $ \;V =3 \quad  $ \\
\hline
  $ \;  8 \quad  $  &  $ 0.10 \quad \;$
  &   $0.29 \quad \;$   & $ 0.53 \quad \; $ \\

 $ \;   16 \quad  $ &  $ 0.24 \quad  \;$ &   $0.76 \quad  \;$
  & $ 1.95 \quad \; $ \\

 $ \; 32 \quad  $  & $ 0.43 \quad  \;$ &   $1.47 \quad  \;$
  & $ 1.71 \quad \; $ \\
\hline \hline
\end{tabular}
\end{table}
The contribution from the charge fluctuations
$\,\widehat{\chi}\sb{cf} \,$ (\ref{57c}) weakly depends on $U$
and $V$ and is very small: $\, \widehat{\chi}\sb{cf}\sim
10^{-3}\,(1/t)\, - 10^{-2}\,(1/t)\, $ for  hole concentrations
$\delta = 0.05 - 0.10$, respectively. For the averaged over the
BZ vertex $ \;\overline{|t({\bf q})|^2} = (1/N)\sum_{\bf
q}|t({\bf q})|^2 \simeq 4\,t^2 $ the contribution induced by the
kinematic interaction is equal to $\overline{|t({\bf
q})|^2}\,\widehat{\chi}\sb{cf} \lesssim 0.04\, t \,$ and can be
neglected.  The charge fluctuation contribution $\,
\widehat{V}_{cf}\,$ (\ref{57b}) from the intersite CI (\ref{50})
for the hole concentration $\delta = 0.05$ is  also small, $\;
\widehat{V}_{cf} \lesssim 5 \cdot 10^{-2}\, t\; $ for $ V \leq 2$
and increases up to $\,0.17 \, t$ for $V=4$. For larger hole
concentration $\, \widehat{V}_{cf}\,$ increases as shown  in
Table \ref{Table1} for $\delta = 0.10$. However,  $\,
\widehat{V}_{cf} - V <0\,$ for all values of   $\,U $ and $\, V$
and consequently,  the $d$-wave pairing induced only by charge
fluctuations cannot occur.

The spin-fluctuation contribution $\, \widehat{\chi}\sb{sf}\,$
(\ref{57e}) is calculated for the model $\chi_{sf}({\bf q})$ in
Eq.~(\ref{51}). Since the spin susceptibility has a maximum at
the AF wave vector ${\bf Q} = (\pi,\pi)$ the integral over ${\bf
k}$ in (\ref{57e}) results in the negative value for $\,
\widehat{\chi}\sb{sf}\,$ which strongly depends on hole doping.
Our previous calculations gave the following values: $\, -
\widehat{\chi}\sb{sf}\cdot t \approx 1.3, \,1.0, \, 0.6$ for hole
concentrations $\delta = 0.05,\, 0.10, \, 0.25$, respectively
(see Ref.~\cite{Plakida13}). Using the averaged over BZ vertex
$\, \overline{|t({\bf q})|^2} \simeq 4\,t^2\,$ we can estimate an
effective spin-fluctuation coupling constant as $\, g_{sf} \simeq
- 4\,t^2 \,\widehat{\chi}\sb{sf} = 5.2,\, 4.0, \, 2.4 $.  Thus,
the spin-fluctuation contribution to the pairing in
Eq.~(\ref{56}) with the coupling constant  $ g_{sf} = 2- 1 $~eV
for $\delta = 0.05 -0.25$  appears to be the largest.
\par
\begin{figure}
\resizebox{0.35\textwidth}{!}{%
\includegraphics{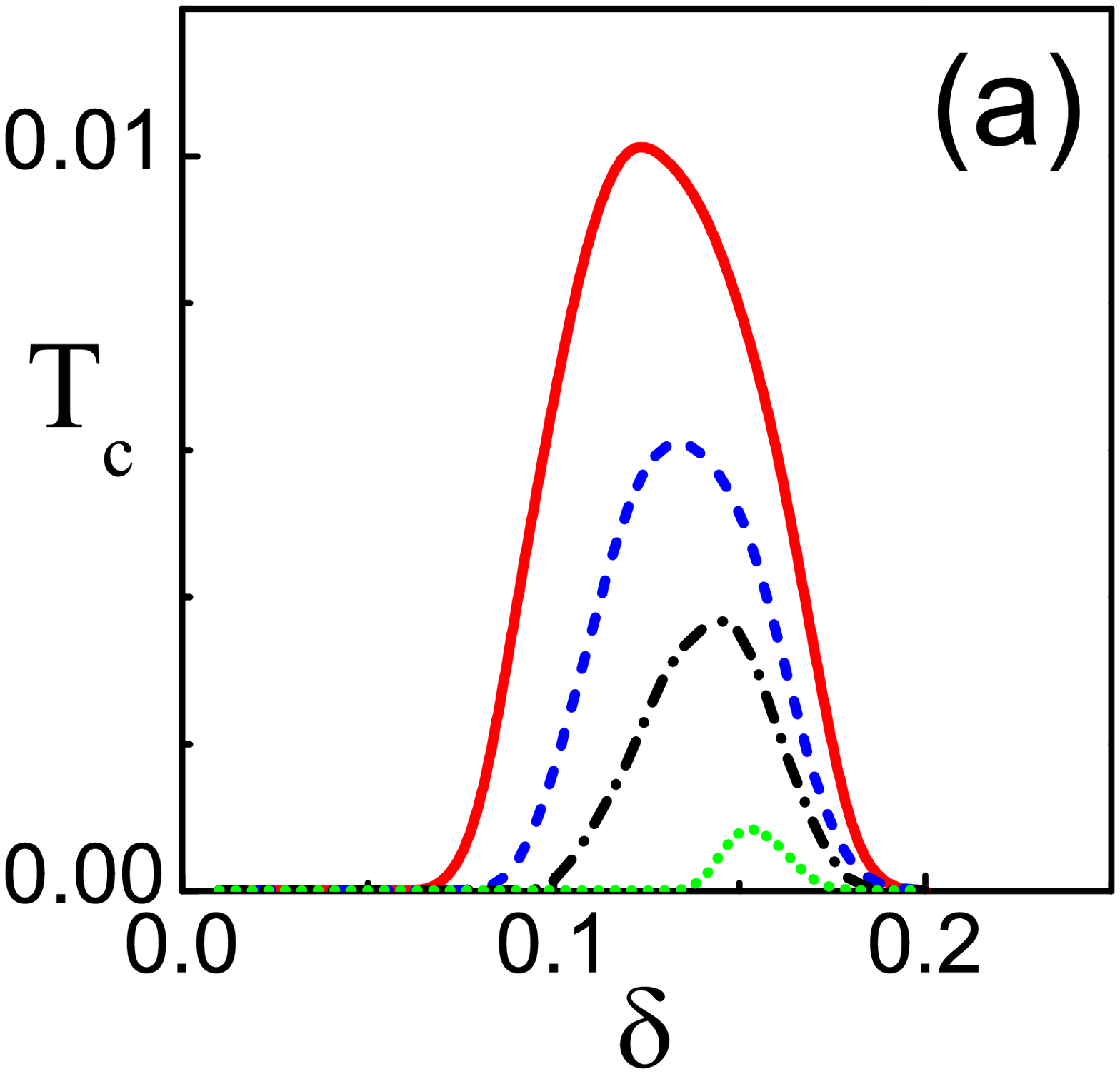}}\vspace{5mm}
\resizebox{0.35\textwidth}{!}{%
\includegraphics{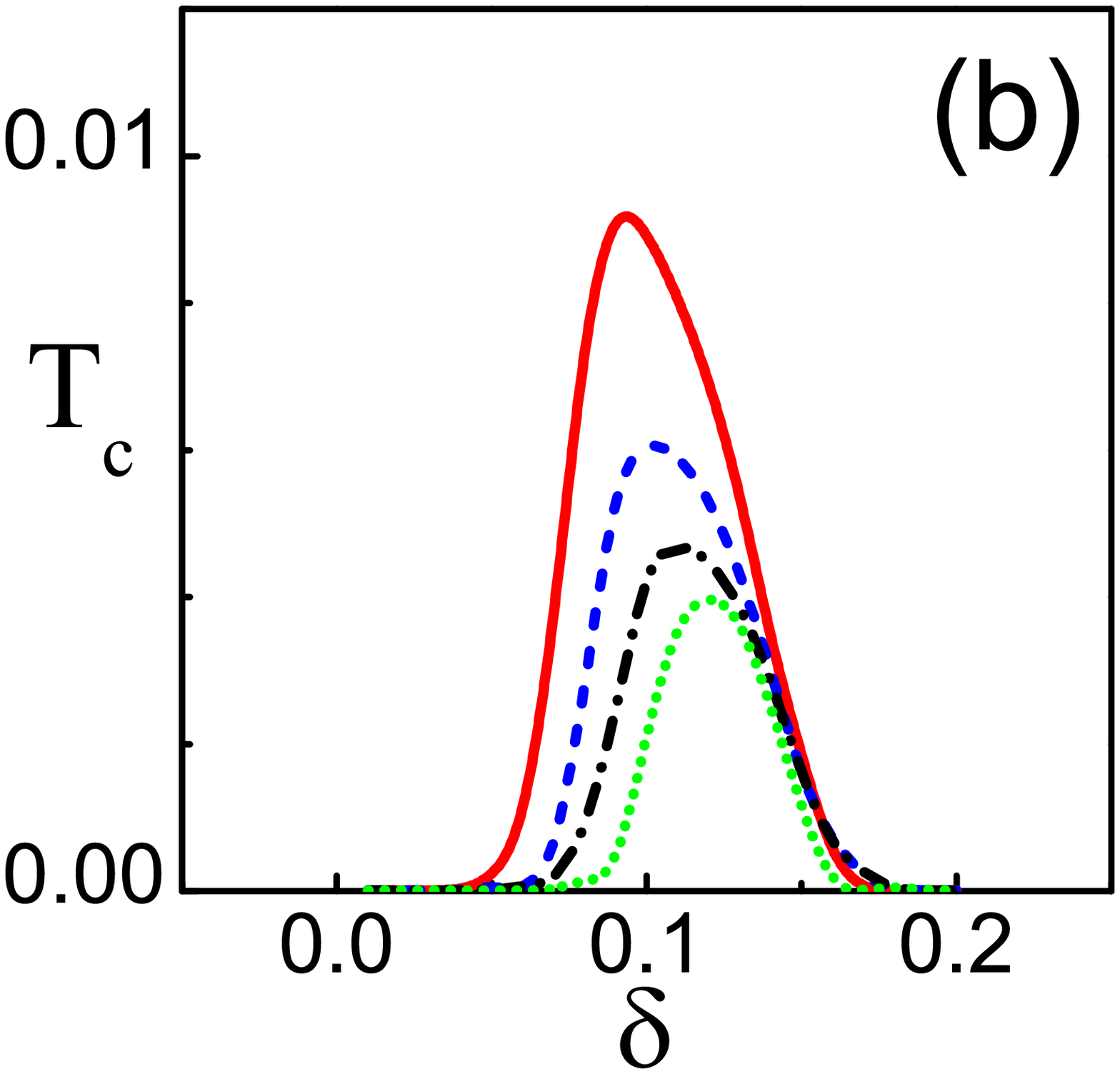}}
\caption{(Color online) $T_c(\delta)$  for  (a) $U = 8 \,$ and
(b) for $U = 16$ for $V = 0.0$  (bold red line), $V = 0.5 $ (blue
dashed line), $V = 1.0 $ (black dash-dotted line), and $V = 2.0 $
(green dotted line). }
 \label{Tc}
\end{figure}
\begin{figure}
\resizebox{0.35\textwidth}{!}{%
\includegraphics{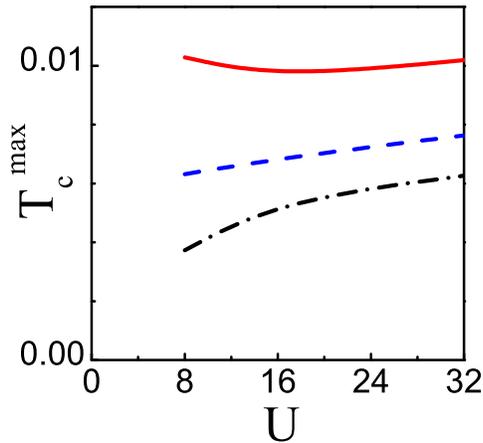}}
 \caption{(Color online)  Maximum $T_c(\delta)$ as a function of  $U \,$
for  $V = 0.0$  (bold red line), $V = 0.5 $ (blue dashed line),
and $V = 1.0 $ (black dash-dotted line).}
 \label{Tc-max}
\end{figure}
Results of  $T_c$ calculation   using   Eq.~(\ref{56}) are shown
in Fig.~\ref{Tc} for (a) $U =8$ and  (b) $U = 16$  and $\, V =
0.0,\, 0.5 ,\,  1.0 $, and $\, 2 \,$. Similar doping dependence
for $T_c$ is observed for  $U = 32$.  The maximum $T_c$ at the
optimal doping as a function of $U$ and  $V$ is shown in
Fig.~\ref{Tc-max}. Increasing of the intersite Coulomb repulsion
$V$ suppresses $T_c$  which becomes small only for high values of
$V = 2t - 3t$ comparable with the spin-fluctuation coupling $\,
g_{sf}$ and much larger than the exchange interaction $\, J = 0.4
t$. At the same time increasing of $U$ enhances $T_c$. This is
due to narrowing of the electronic band as seen in
Figs.~\ref{Ek8},~\ref{Ek16} and corresponding increase of the
density of state.
\begin{figure}[ht!]
\resizebox{0.35\textwidth}{!}{%
\includegraphics{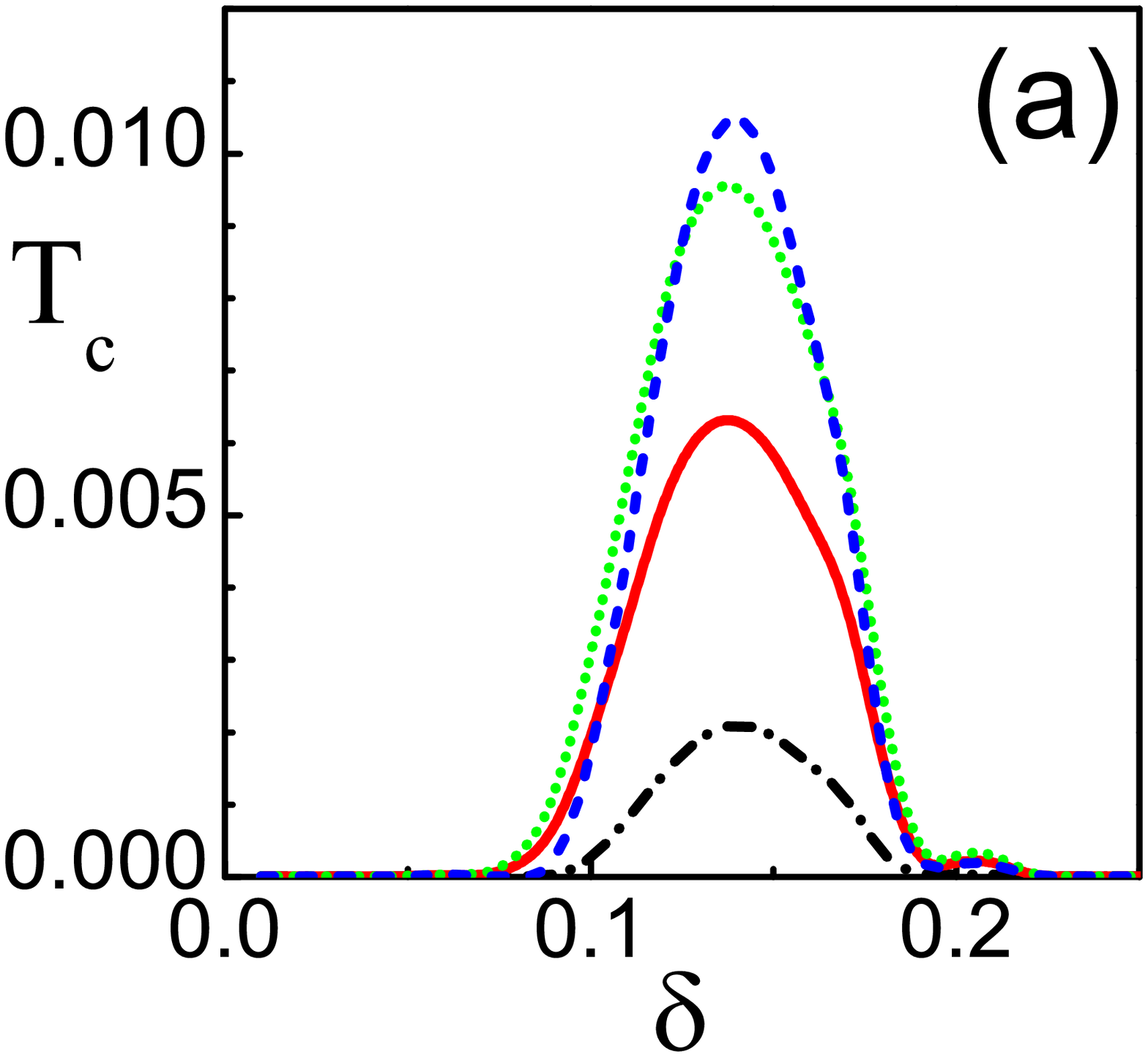}}
\resizebox{0.35\textwidth}{!}{%
\includegraphics{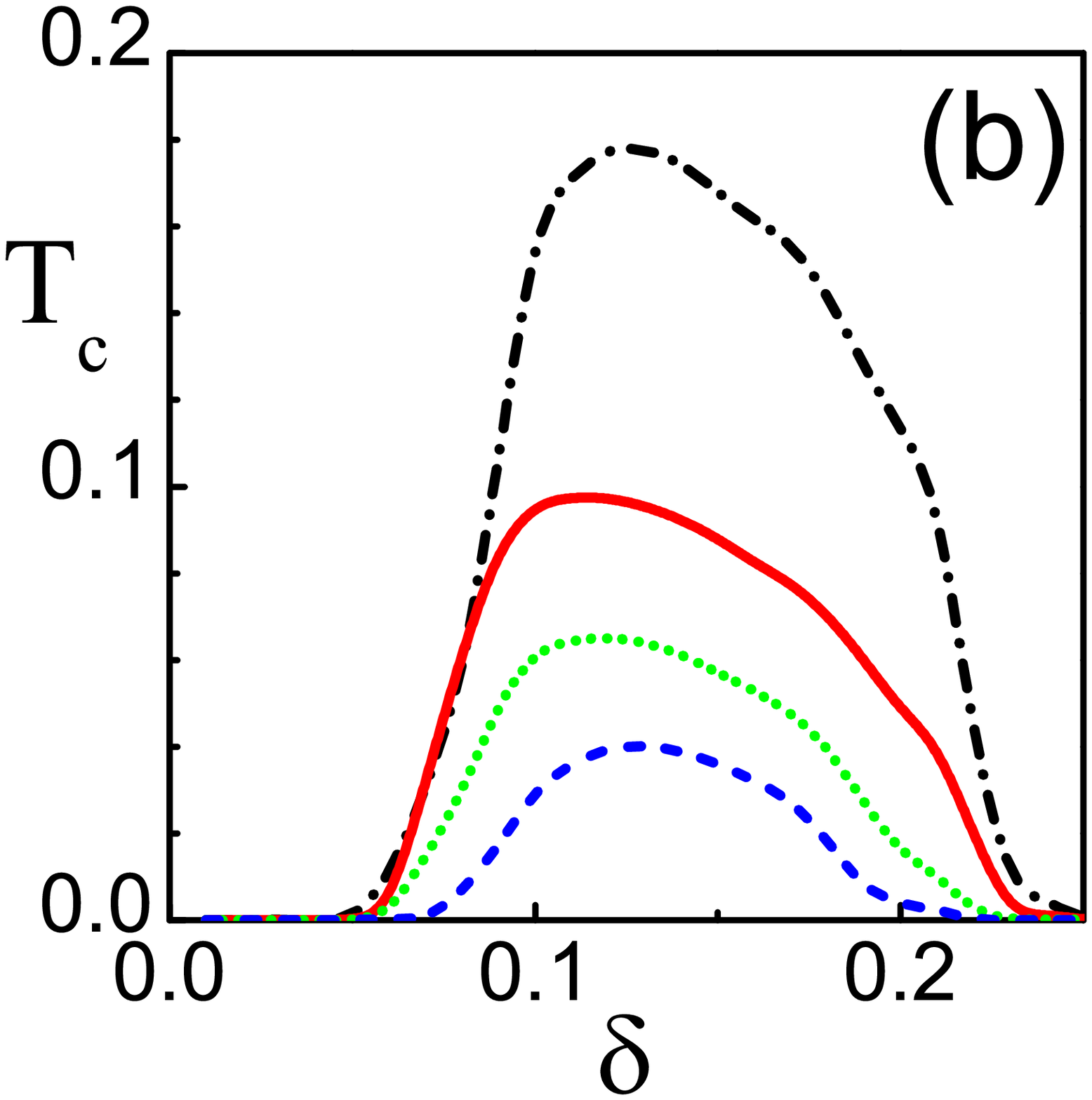}}
\caption{(Color online)  $T_c(\delta)$ dependence on
spin-fluctuation contribution $\chi_{sf}$ in Eq.~(\ref{56})  for
$\omega_s = 0.2$ (black dash-dotted line), $\omega_s = 0.4$ (bold
red line) , $\omega_s = 0.6$ (green dotted line), and  $\omega_s
=1.0$ (blue dashed line) calculated for (a) finite $Z({\bf q})$
and    (b) $Z({\bf q}) = 1$. }
 \label{Tc-oms}
\end{figure}
\begin{figure}
\resizebox{0.35\textwidth}{!}{%
\includegraphics{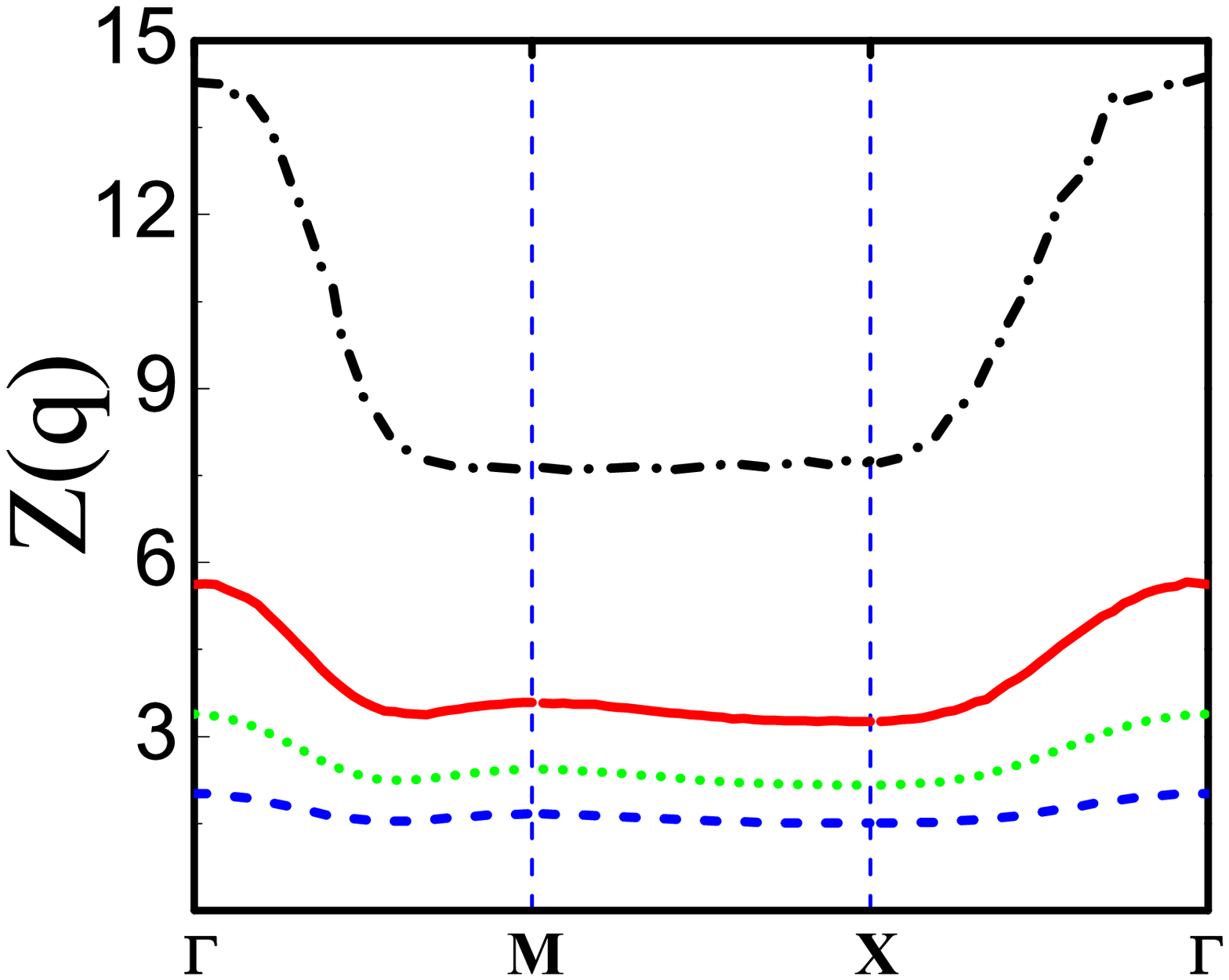}}
 \caption{(Color online) $Z({\bf q})
$  dependence on  spin-fluctuation contribution $\chi_{sf}$ in
Eq.~(\ref{45ba}) for $\omega_s = 0.2$ (black dash-dotted line),
$\omega_s = 0.4$ (bold red line) , $\omega_s = 0.6$ (green dotted
line), and $\omega_s =1.0$ (blue dashed line) at $\delta = 0.10$.}
 \label{Zq-oms}
\end{figure}

To prove  an important role of the spin-fluctuation interaction
both in the normal state and in superconducting pairing we
calculate the function  $Z({\bf q})$ (\ref{45ba}) and   $T_c$  for
several values of the parameter $\omega_s \,$ for the static
susceptibility in the model (\ref{51}): $\,\omega_s = 0.2 , \,
0.4,\,  0.6\, $ and $1.0 \,$ for $U = 8 \,$. Figure~\ref{Tc-oms}
shows $T_c$ dependence on the parameter $\omega_s$ that
determines  the spin-fluctuation contribution
$\widehat{\chi}\sb{sf}$ in Eq.~(\ref{56}) in two cases: for (a)
$Z({\bf q})$ given by Eq.~(\ref{45ba}) and (b) $Z({\bf q}) = 1$.
Since the spin-fluctuation interaction is determined by $ \chi_{
Q} \propto 1/\omega_{s}$ (\ref{52}) it increases  with  lowering
of the cut-off frequency $\, \omega_s $.   This results in
increasing of the superconducting pairing contribution
$\widehat{\chi}\sb{sf}$ but at the same time enhances the normal
state renormalization $Z({\bf q})$ as shown in Fig.~\ref{Zq-oms}.
Therefore, in the case (a) $T_c$, roughly being proportional to
$\widehat{\chi}\sb{sf}/[Z({\bf q})]^2 $, decreases due to
suppression of the QP weight $1/Z({\bf q})$, while in the case (b)
for $Z(\bf q) =1$ increasing of pairing strength results in $T_c$
increase. Note also, that $T_c$ in Fig.~\ref{Tc-oms} (b)
calculated in MFA with $Z({\bf q}) =1$ an order of magnitude
larger than its value with a proper consideration of electronic
spectrum renormalization.
\par
In the current approach one can also consider  the $s$-wave
pairing. For the extended $s$-wave gap function, $\varphi_s({\bf
k}) = (\Delta/2) \,\eta_s({\bf k})$ where $ \eta_s({\bf k}) =
(\cos k_x + \cos k_y)$, a similar to (\ref{55}) equation for
$T_c$ can be derived. Solution of this equation reveals a finite
and quite high $T_c$. However, $s$-wave pairing symmetry violates
a kinematic restriction of no double occupancy for the Hubbard
model in the two-subband regime. As was pointed out in
Refs.~\cite{Plakida89,Yushankhai91}, the single-site correlation
function should obey the condition
\begin{equation}
\langle X_{i}^{\bar\sigma 2} X_{i}^{\sigma2} \rangle =
 \frac{1}{N} \sum_{\bf q}\,
\langle X_{-\bf q}^{\bar\sigma 2} X_{\bf q}^{\sigma2} \rangle =0 ,
 \label{w1}
\end{equation}
caused by the multiplication rule for the Hubbard operators, $\,
X_{i}^{\alpha\beta} X_{i}^{\gamma\delta} = \delta_{\beta\gamma}
X_{i}^{\alpha\delta} \,$. In the QP approximation used in
Eq.~(\ref{55}) we obtain the relation
\begin{equation}
\langle X_{i}^{\bar\sigma 2} X_{i}^{\sigma2} \rangle =
 \frac{1}{N} \sum_{\bf q}\, \frac{\varphi(\bf q)}
  {[Z({\bf q})]^2 \; 2\widetilde{\varepsilon}({\bf q})}
    \tanh\frac{\widetilde{\varepsilon}({\bf q})}{2T_c} = 0.
     \label{w2}
\end{equation}
For the $d$-wave pairing $\varphi_d({\bf q}) = (\Delta/2) \,(\cos
q_x - \cos q_y)$ this condition is  fulfilled in the tetragonal
phase for any doping (pairing in the orthorhombic pase is
considered in Ref.~\cite{Plakida00}). For the $s$-wave pairing
this condition is violated
\begin{equation}
 \frac{1}{N} \sum_{q_x , q_y}\, \frac{\cos q_x}
  {[Z({\bf q})]^2 \; 2\widetilde{\varepsilon}({\bf q})}
    \tanh\frac{\widetilde{\varepsilon}({\bf q})}{2T} \neq 0,
     \label{w3}
\end{equation}
for an arbitrary doping except for a particular choice  of the
chemical potential when the contribution from the integral over
$0 \leq q_x \leq \pi$ is compensated  by the integral over $\pi
\leq q_x \leq 2\pi$.  The same condition holds  for the
one-particle subband, $\, \langle X_{i}^{0 \bar\sigma} X_{i}^{0
\sigma} \rangle =0\,$. The obtained results can be derived for a
general representation for the correlation function
\begin{eqnarray}
\langle X_{-\bf q}^{\bar\sigma 2} X_{\bf q}^{\sigma2} \rangle  =
 -\frac{1}{\pi Q_2 N}  \sum_{\bf q}
  \int\limits\sb{-\infty}\sp{+\infty}\frac{ dz}
   {{\rm e}^{z/T}  +1}
  \mbox{Im} \, F\sb{\sigma}\sp{22}({\bf q},z),
 \nonumber
\end{eqnarray}
since the  symmetry of the  anomalous GF $
F\sb{\sigma}\sp{22}({\bf q},z)\,$ is determined by the $s$-  or
$d$-wave symmetry of the gap function. Therefore, we conclude
that $s$-wave pairing is prohibited for the Hubbard model in the
limit of strong correlations.

\subsection{Comparison with previous theoretical studies}
\label{sec:10}

As  discussed in Sec.~\ref{sec:1}, the intersite Coulomb
repulsion $V$ is detrimental for pairing induced by the on-site
CI $U$ in the Hubbard model or higher-order contributions from
$V$  in the weak correlation limit. Here we would like to comment
on several studies of this problem  in the strong correlation
limit and to compare them with our analytical results for the
$d$-wave pairing.

Following the original idea of Anderson~\cite{Anderson87}, it is
commonly believed that the exchange interaction $\,J = 4t^2/U \,$
induced by the interband hopping in the Hubbard model plays the
major role in the $d$-wave superconducting pairing. Since the
excitation energy of electrons in the interband hopping $\, U\, $
is much larger than their intraband kinetic energy $\,  W \,$ the
exchange pairing has no retardation effects contrary to the
electron-phonon pairing where large Bogoliubov-Tolmachev
logarithm~\cite{Bogoliubov58} diminishes the Coulomb repulsion $V
\rightarrow V/[1 + \rho_c\, \ln(\mu/\omega_{ph})]$ where $\rho_c
= N(0)\, V $ and $\omega_{ph}$ is the phonon energy.
Consequently, without the retardation effects the  Coulomb
repulsion $V$ should destroy the exchange  pairing for $\,V >
J\,$.
\par
To get over this problem,  in Ref.~\cite{Senechal12} it was
suggested that in the limit of strong correlation the intersite
Coulomb repulsion $V$  decreases the interband excitation energy
which results in enhancement of the exchange interaction,  $\,
\widetilde{J}(V) = {4\, t^2}/(U-V) \,$, as was found from cluster
calculations. If we consider pairing induced only by the exchange
interaction $\widetilde{J}(V)$ and take into account the Coulomb
repulsion $V$ then  the condition $ \, \widetilde{J}(V) - V >
0\,$ should be fulfilled for existence of pairing.  The condition
is satisfied  for $\,0 < V < V_1$ where $\, V_{1} = (U/2)[ 1-
\sqrt{1 - (4\, t/U)^2} \;]$  for $\,0 \leq V < U $. For $U > 4t$
we have $V_1 \ll U$ as, e.g.,  for $\, U =8 $,  $\, V_1 = 0.067
\, U \,$ and for $\, U = 32 $, $\, V_1  = 0.004 \,  U $.
Therefore, we see that the pure exchange superconducting pairing
can occur in the region of weak Coulomb repulsion. Contrary to
this,  in Ref.~\cite{Senechal12} using the cellular dynamical
mean-Field theory (CDMFT)~\cite{Senechal12r} the $d$-wave pairing
was found in the region of strong coupling up to $V \lesssim U/2
$ (as, e.g., shown in Fig.~3, $\, V \leqslant 3\, t \,( 8\, t)$
for $U = 8\, t \,( 16\, t)$, respectively). At the same time, in
the limit of weak correlations $U =4t\,$  the pairing is
suppressed at the smaller value of  $V \sim 1.5\, t \,$. Thus, we
believe that ``Resilience of $d$-wave superconductivity to
nearest-neighbor repulsion'' is not due to renormalization of the
exchange interaction $\widetilde{J}(V)$ but due to another
mechanism of pairing not explicitly seen in the CDMFT
calculations. As we have shown in the strong correlation limit in
the two-subband regime the emerging kinematic interaction is
responsible for the spin-fluctuation pairing at large values of
$V$, up to $V \lesssim 4\, t$.
\par
Our  conclusion about  importance of the kinematic mechanism of
pairing is supported by the studies in Ref.~\cite{Plekhanov03}.
Using the variational Monte Carlo technique the superconducting
$d$-wave  gap was calculated for  the extended Hubbard model with
a weak exchange interaction $J = 0.2 \, t$  and a repulsion $V
\leq 3\, t $ in a broad range of $\,0 \leq U \leq 32$. It was
found that the gap decreases with increasing $V$ at all $U$ and
can be suppressed for  $V > J $ for small $U $. But for large $
\,U \gtrsim U_c \sim 6\, t\,$ the gap becomes  robust and exists
up to  large values of $V \sim 10\, J = 2\, t $ which was
explained by effective enhancement of   $J$ as in
Ref.~\cite{Senechal12}. At the same time, the gap does not show
notable variation with $U$ for large $U = 10 - 30$ though it
should depend  on the conventional exchange interaction in the
Hubbard model $J = 4t^2/U$ (or $J = 4t^2/(U - V) $). We can
suggest another explanation of these results by pointing out that
at large $U \gtrsim U_c \,$ concomitant decrease of the bandwidth
(as shown in Fig. 3~b) in Ref.~\cite{Plekhanov03}) results in the
splitting of the Hubbard band into the upper and lower subbands
and the emerging kinematic interaction induces the $d$-wave
pairing in one Hubbard subband. In that case the second subband
for large $U$ gives a small contribution which results in
$U$-independent pairing. It can be suppressed by the repulsion
$V$ only larger than the kinematic interaction, $V \gtrsim 4t$.
\par
In Ref.~\cite{Raghu12a} the extended Hubbard model is considered
in the weak or intermediate correlation limits as in
Ref.~\cite{Raghu12} and in the strong correlation limit within
the slave-boson representation in the mean-field approximation
(MFA). In the strong correlation limit a small value of $V = J$
was found which suppresses the $d$-wave superconducting gap.
However, in the MFA  the kinetic energy term described by the
projected electron operators, $t\, \hat{c}_{i\sigma }^\dag
\hat{c}_{j \sigma } = t\, {c}_{i\sigma }^\dag (1 -  n_{i - \sigma
}){c}_{j \sigma }(1 - n_{j - \sigma })\equiv t\, X_{i}^{\sigma 0}
X_{j}^{ 0\sigma} $, is approximated by the conventional fermion
(spinon) operators, $t\,\delta \, f_{i\sigma }^\dag f_{j \sigma }
$ and the most important contribution from the kinematic
interaction  is lost in the resulting BCS-type gap equation (13)
in Ref.~\cite{Raghu12a}. As shown in our equation for the gap
(\ref{56}) the kinematic interaction given by
$\widehat{\chi}\sb{sf}$ (\ref{57e}) provides strong
spin-fluctuation pairing and high $T_c$.
\par
To analyze the pairing mechanisms in the limit of strong
correlations analytical methods should be used.  A complicated
dynamics of projected electron operators can be rigorously taken
into account using the HO technique. The algebra of the HOs
preserves rigorously restriction of no double occupancy of
quantum states which is violated in the commonly used MFA in the
slave-particle theory. As discussed in Sec.~\ref{sec:2a}, the
commutation relations for the HOs results in the kinematic
interaction which is responsible for strong spin-fluctuation
electron interaction. The superconducting pairing induced by the
kinematic interaction for the HOs was first proposed  by Zaitsev
and Ivanov~\cite{Zaitsev87} who studied the two-particle vertex
equation by applying the diagram technique for HOs. The
momentum-independent $s$-wave superconducting gap was found
which, however,  violates the HO kinematics  as was shown in
Refs.~\cite{Plakida89,Yushankhai91} (see Eqs.~(\ref{w1}) --
(\ref{w3})).  Since the intersite Coulomb repulsion $V > J$
destroys the superconductivity induced by the AF exchange
interaction, the spin-fluctuation pairing in the second order of
the kinematic interaction  beyond the GMFA should be taken into
account  as discussed in detail in Sec.~\ref{sec:9} and  for the
$t$--$J$ model was considered  in
Refs.~\cite{Plakida99,Prelovsek05}.

\section{Conclusion}
\label{sec:11}

In the paper we have studied effects of the strong intersite
Coulomb repulsion $V$ on the $d$-wave  superconducting pairing
within the extended Hubbard model (\ref{1}) in the limit of
strong electron correlations, $U \gg t$. Using the Mori-type
projection technique we  obtained a self-consistent system of
equations for normal and anomalous (pair) GFs and for the
self-energy calculated in the SCBA.
\par
It was found that the kinematic spin-fluctuation interaction
$g_{sf}$ induced by electron hopping in one Hubbard subband is
much stronger than the conventional exchange interaction $J$
resulting from the interband hopping. Consequently, the  $d$-wave
pairing can be suppressed only for large values of $V > g_{sf}$
where $g_{sf}\,$ is of the  order of kinetic energy $g_{sf}\sim W
\approx 4t\,$. Since in the cuprates the Coulomb repulsion $V$ is
of the same order  as the exchange interaction, $\, V  \gtrsim J
\sim 0.4\,t $, the kinematic spin-fluctuation  pairing mechanism
plays the major role in achieving high-temperature
superconductivity. It is also shown that the kinematic
spin-fluctuation interaction results in a strong renormalization
of electronic spectra.
\par
It is important to point out that the superconducting pairing
induced by the AF exchange interaction and the spin-fluctuation
kinematic interaction is characteristic for systems with strong
electron correlations. These mechanisms of superconducting
pairing are absent in the fermionic models  and are generic for
cuprates. Therefore, we believe that the spin-fluctuation
kinematic mechanism of superconducting pairing in the Hubbard
model in the limit of strong correlations is the relevant
mechanism of high-temperature superconductivity in the
copper-oxide materials.

\acknowledgments

The authors would like to thank A.S. Alexandrov, V.V. Kabanov,
A.-M. S. Tremblay and M.Yu. Kagan for valuable discussions.
Partial financial support by the Heisenberg--Landau Program of
JINR is acknowledged.


\begin{thebibliography}{99}
\bibitem{Schrieffer07}  \textit{ Handbook of High-Temperature  Superconductivity.
Theory and Experiment}, edited by J.\ R.~Schrieffer  and J.\
S.~Brooks
(Springer-Verlag, New York, 2007).
\bibitem{Plakida10}  N.\ M. Plakida,
\textit{ High-Temperature Cuprate Superconductors} (Springer
Series in Solid-State Sciences, Vol. 166, Springer-Verlag,
Berlin, 2010), Chap. 7.
\bibitem{Anderson87}  P.\ W.~Anderson,  Science
\textbf{235},  1196 (1987);  P.\ W.~Anderson, \textit{The theory
of superconductivity in the high-$T\sb{c}$ cuprates} (Princeton
University Press, Princeton, 1997).
\bibitem{Hubbard63} J. Hubbard, Proc. Roy. Soc. (London)
A,  \textbf{276}, (1963) 238.
\bibitem{Alexandrov11} A.\ S. Alexandrov and V.\ V. Kabanov,
Phys. Rev. Lett. \textbf{106}, 136403 (2011).
\bibitem{Raghu12} S. Raghu,   E. Berg, A.\ V. Chubukov,
and S.\ A. Kivelson  Phys. Rev. B {\bf 85}, 024516 (2012).
\bibitem{Kohn65} W. Kohn and J.\ M. Luttinger, Phys. Rev. Lett.  \textbf{15}, 524
(1965).
\bibitem{KaganM11} M.\ Yu. Kagan, D.\ V. Efremov, M.\ S. Marienko, and V.\ S. Val'kov.
JETP Lett. \textbf{93} 725 (2011).
\bibitem{Efremov00} D.\ V. Efremov, M.\ S. Mar'enko, M.\ A. Baranov, and M.\ Yu. Kagan,
J. Exp. Theor. Phys.  \textbf{90}, 861 (2000).
\bibitem{KaganM13}  M.\ Yu. Kagan, V.\ V. Val'kov, V.\ A. Mitskan, and M.\ M. Korovushkin,
J. Exp. Theor. Phys.  \textbf{144}, 837 (2013).
\bibitem{Dagotto94} E. Dagotto, Rev. Mod. Phys. \textbf{66},
 763 (1994).
\bibitem{Bulut02} N. Bulut, { Advances in Physics}
 \textbf{51}, 1587 (2002).
\bibitem{Scalapino07}  D.\ J. Scalapino, Numerical  studies  of the 2D Hubbard model,
in Ref.~\cite{Schrieffer07}, pp. 495--526.
\bibitem{Senechal12r}  D. S\'{e}n\'{e}chal, Cluster dynamical
mean field theory, in {\it Theoretical methods for Strongly
Correlated Systems,} edited by A. Avella and F. Mancini (
Springer Series in Solid-State Sciences, Vol. 171, Springer-
Verlag, Berlin, 2012), Chap. 11.
\bibitem{Plekhanov03} E. Plekhanov, S. Sorella, and M. Fabrizio,
 Phys. Rev. Lett. {\bf 90}, 187004  (2003).
\bibitem{Senechal12}  D. S\'{e}n\'{e}chal, A. Day, V. Bouliane, and A.-M.\ S.
Tremblay, arXiv:1212.4503 [cond-mat.supr-con].
\bibitem{Raghu12a} S. Raghu, R. Thomale, and T.\ H. Geballe, Phys. Rev. B  {\bf
86}, 094506 (2012).
\bibitem{Plakida13}   N.\ M.~Plakida  and  V.\ S.~Oudovenko,
Eur. Phys. J. B \textbf{86}, 115 (2013).
\bibitem{Hubbard65} J. Hubbard,
Proc.~Roy. Soc. A (London) \textbf{285},  (1965) 542.
\bibitem{Mori65} H. Mori, Prog. Theor. Phys. \textbf{34},
399 (1965).
\bibitem{Zubarev60} D.\ N. Zubarev,
{ Usp. Fiz. Nauk}  \textbf{71}, 71  (1960); (Sov. Phys. Usp.
\textbf{3}, 320 (1960)); {\it Nonequilibrium Statical
Thermodynamics} (Consultant Bureau, New-York, 1974).
\bibitem{Plakida07}   N.\ M.~Plakida  and  V.\ S.~Oudovenko,
JETP \textbf{104}, 230 (2007).
\bibitem{Plakida03} N.\ M. Plakida,   L.~Anton, S.~Adam,
and Gh.~Adam, Zh. Exp.Theor. Fyz. \textbf{124}, 367 (2003), (JETP
\textbf{97}, 331 (2003)).
\bibitem{Eliashberg60} G.\ M.~Eliashberg, Zh. Eksp. Teor. Fiz. \textbf{38}, 966 (1960);
ibid  \textbf{39}, 1437 (1960) (Soviet Phys.~JETP \textbf{11},
696 (1960); ibid \textbf{12},  1000 (1960)).
\bibitem{Jaklic95} J. Jakli\v{c} and P. Prelov\'sek,
 Phys. Rev. Lett. \textbf{74}, 3411 (1995);
{ ibid.} \textbf{75}, 1340 (1995).
\bibitem{Vladimirov09} A.\ A. Vladimirov, D. Ihle, and N.\ M. Plakida,
Phys. Rev. B \textbf{80}, 104425 (2009).
\bibitem{Bogoliubov58}
N.\ N. Bogoliubov, V.\ V. Tolmachev,  and D.\ V. Shirkov,   {\it
New method in the theory of superconductivity},    (Publ. Dept.
USSR Acad. of Science, Moscow, 1958; Consultants Bureau,  Chapman
and Hall, New York - London, 1959, Vol. YII).
\bibitem{Zaitsev87} R.\ O. Zaitsev,  and  V.\ A. Ivanov,
{Soviet Phys. Solid State} \textbf{29},    2554  (1987),
{ Ibid.} {\bf 29},   3111 (1987), {Int. J. Mod. Phys. B}
\textbf{5},  153  (1988).
\bibitem{Plakida89} N.\ M. Plakida, V.\ Yu. Yushankhai, and
I.\ V. Stasyuk, { Physica C}  \textbf{160}, 80 (1989).
\bibitem{Yushankhai91} V.\ Yu. Yushankhai, N.\ M. Plakida,  and P. Kalinay,
{ Physica C} \textbf{174}, 401 (1991).
\bibitem{Plakida00}  N.\ M.~Plakida  and  V.\ S.~Oudovenko,
 Physica C, {\bf 341--348}, 289 (2000);
in {\it Proceedings of the NATO ARW on open problems in strongly
correlated systems}, edited by  J.Bon·ca, P. Prelov·sek, A.
Ram·sak and S. Sarkar (Kluewer Academic Publs., 2001) p. 111-116.
\bibitem{Plakida99}   N.\ M.~Plakida  and  V.\ S.~Oudovenko,
 { Phys.~Rev.~B}  \textbf{59}, 11949 (1999).
\bibitem{Prelovsek05}  P. Prelov\v{s}ek and A. Ram\v{s}ak,
 Phys.~Rev.~B {\bf 72}, 012510 (2005).
\end{thebibliography}
\end{document}